\documentclass[fleqn,11pt]{wlscirep}
\usepackage[utf8]{inputenc}
\usepackage[T1]{fontenc}

\title{Harder, shorter, sharper, forward:\\A comparison of women's and men's elite football gameplay (2020-2025)}

\author[1]{Rebecca~Carstens}
\author[2]{Raj~Deshpande}
\author[3]{Pau~Esteve}
\author[4]{Nicolò~Fidelibus}
\author[5]{Sara~Linde~Neven}
\author[6]{Ramona~Ottow}
\author[7]{Lokamruth~K.~R.}
\author[8]{Paula~Rodríguez-Sánchez}
\author[9,10]{Luca~Santagata}
\author[8]{Javier~M.~Buldú}
\author[10,11,12]{Brennan~Klein}
\author[10,*]{Maddalena~Torricelli}

\affil[1]{Roslin Institute, University of Edinburgh, Edinburgh, United Kingdom}
\affil[2]{NPLab, Network Science Institute, Northeastern University London, London, United Kingdom}
\affil[3]{Instituto de Física Interdisciplinar y Sistemas Complejos IFISC, Campus UIB, Palma de Mallorca, Spain}
\affil[4]{Department of Science, Technology and Society, Scuola Universitaria Superiore IUSS Pavia, Pavia, Italy}
\affil[5]{Institute for Biodiversity and Ecosystem Dynamics, University of Amsterdam, Amsterdam, Netherlands}
\affil[6]{Universitat Pompeu Fabra, Barcelona, Spain}
\affil[7]{Department of Computer Science, Reykjavík University, Reykjavík, Iceland}
\affil[8]{Complex Systems Group, Universidad Rey Juan Carlos, Móstoles, Madrid, Spain}
\affil[9]{Information Engineering and Computer Science Department, University of Trento, Povo, Italy}
\affil[10]{Network Science Institute, Northeastern University, Boston, Massachusetts, USA}
\affil[11]{Department of Physics, Northeastern University, Boston, Massachusetts, USA}
\affil[12]{Institute for Experiential AI, Northeastern University, Boston, Massachusetts, USA}

\affil[*]{Corresponding author: m.torricelli@northeastern.edu}

\begin{abstract}
Elite football is believed to have evolved in recent years, yet systematic evidence for the pace and form of that change remains sparse. Drawing on event-level records for 13,018 matches across ten top-tier men's and women's leagues in England, Spain, Germany, Italy, and the United States (2020-2025), we quantify match dynamics through two complementary lenses: conventional performance statistics and pitch-passing networks that track ball movement across spatial regions of the field. Between 2020 and 2025, average passing volume, pass accuracy, and the proportion of passes made under pressure all increased, with the largest year-on-year changes occurring in women's competitions. Network measures reveal that normalized outreach decreased, indicating teams increasingly concentrate ball circulation into shorter-range passing connections rather than wide spatial distribution. These trends are consistent across countries and tiers, yet persistent national differences indicate that stylistic diversity remains. Notably, women's competitions exhibit stronger rates of change across most metrics, consistent with an accelerating professionalization, while the systematic decline in network outreach across all competitions points to a sport-wide tactical convergence toward shorter, more concentrated passing structures.
\end{abstract}

\begin{document}

\flushbottom
\maketitle

\thispagestyle{empty}

\section*{Introduction}

Football has undergone profound transformations throughout its history, shaped by rule changes, technological advances, and shifting social contexts. The sport's professionalization began in the late 19th century with the establishment of the English Football League in 1888 \cite{williams2003}, yet women's football faced decades of marginalization---including a ban on women's matches at affiliated pitches in England from 1921 to 1971 \cite{lopez1997}. In recent years, the landscape has become increasingly dynamic: new teams join top-tier leagues; competitions expand and restructure; technological innovations such as the Video Assistant Referee (VAR) alter gameplay; and women's leagues experience unprecedented growth in visibility, investment, and institutional support \cite{fifa2023, deloitte2025wsl}. This convergence of structural shifts, regulatory evolution, and cultural change makes contemporary football an inherently complex and rapidly evolving system.

The abundance of detailed match data in recent years has fueled the growth of sports analytics as a vibrant interdisciplinary field. Network science has been applied to analyze dynamics in various sports---including football \cite{duch2010}, basketball \cite{fewell2012}, boxing \cite{tennant2020}, badminton \cite{gomez2020}, rugby \cite{cintia2015}, and water polo \cite{passos2011}---with football being the most extensively studied. Different network representations have been employed to describe connections between players, teams, or leagues \cite{ribeiro2017,buldu2018,ramos2018,pan2025,pena2012,li2025,aguirre2013}. Match outcome data have been used to construct club networks assessing competitive influence \cite{basini2023}; event data have enabled identification of key players and analysis of player interactions \cite{pappalardo2018, pappalardo2019playerank, pappalardo2019public}; and tracking data have supported studies of player coordination at fine spatial scales \cite{buldu2020, buldu2019}. Advanced machine-learning methods now enable in-depth match analysis \cite{spearman2017, fernandez2018, rico2023, biermann2023, seckin2023}, while research on tactical evolution documents changing playing styles across leagues \cite{gonzalezrodenas2023tactics, gonzalezrodenas2024tactical, bauer2023, wilson2008, goldblatt2014}. Recent work has also examined the temporal dynamics of goal scoring and momentum effects \cite{ayana2025}.

Despite this rich body of work, systematic comparative analyses tracking the evolution of tactical and strategic aspects of gameplay across both men's and women's competitions remain sparse. National football cultures shape how the game is played \cite{wilson2008}: Spain is known for technical, possession-heavy football; Germany emphasizes structured organization and high pressing; Italy prioritizes tactical discipline and defensive solidity; England blends physicality with speed; and the United States focuses on athleticism and direct attacking play. Structural differences reinforce these tendencies: the English Premier League is highly commercialized with foreign ownership \cite{plnow2025}; Spain and Italy also allow foreign ownership but with variable financial stability \cite{ft2025}; Germany's Bundesliga enforces the ``50+1'' rule ensuring fan influence \cite{bauers2019}; while Major League Soccer follows a franchise-based model with centralized ownership and no promotion/relegation \cite{guardian2025}. Women's leagues in these countries often share governance structures with their male counterparts \cite{weatherillownership}, yet playing styles are less distinctly defined due to ongoing professionalization and greater fluidity in player movement and coaching approaches \cite{pappalardo2021}. Beyond structural and cultural differences, football's evolution is shaped by season-to-season variability driven by player transfers, coaching changes, and external events. Major tournaments such as the 2022 FIFA World Cup and 2024 European Championship impact subsequent season starts as players require recovery time, affecting early-season performance patterns \cite{molango2025}. Within seasons, short-term dynamics including player fatigue, injuries, and team form influence match-level metrics. These temporal factors introduce complexity into evolutionary trend analysis, as improvements in some metrics may reflect adaptation to coaching systems while declines in others may signal accumulated physical strain.

In this study, we investigate the evolution of football gameplay from a tactical and strategic perspective, analyzing men's and women's elite competitions in parallel across recent seasons. Our central question is: How have key performance indicators evolved across top-tier leagues between 2020 and 2025, and do these changes differ systematically across countries and between men's and women's football?

We analyze 13,018 matches from ten competitions across five countries (England, Spain, Germany, Italy, and the United States) spanning 2020 to 2025 \cite{statsbomb_private_2025}. Our approach integrates two complementary lenses. First, we employ statistical analysis of team-level match performance indicators---such as passing volume, accuracy, and pressure---to quantify observable changes in gameplay. Second, we construct pitch-passing networks where spatial zones of the pitch serve as nodes connected by passes, allowing us to capture ball possession dynamics and spatial control \cite{zhou2023, herrera2020, garrido2020, novillo2024, huang2024}. By focusing on team-level attributes rather than individual players, we capture broader structural trends in playing style and match behavior.

Our analysis reveals systematic patterns in how football is played and how these patterns have changed over the past five seasons. We find increases in possession intensity and passes made under defensive pressure, particularly pronounced in European leagues, with women's competitions exhibiting stronger rates of change across most metrics. Spatial organization shows divergent trends: declining network outreach and increasing vertical play suggest a shift toward more compact, direct attacking patterns. Country-level differences persist throughout, with Spain maintaining distinct tactical signatures in both offside rates and spatial distribution. Finally, we identify league-specific evolutionary trajectories---most notably in Italy's Serie A Women, which shows rapid tactical modernization distinct from other competitions.

\section*{Results}

\begin{figure}[ht!]
\centering
\includegraphics[width=\linewidth]{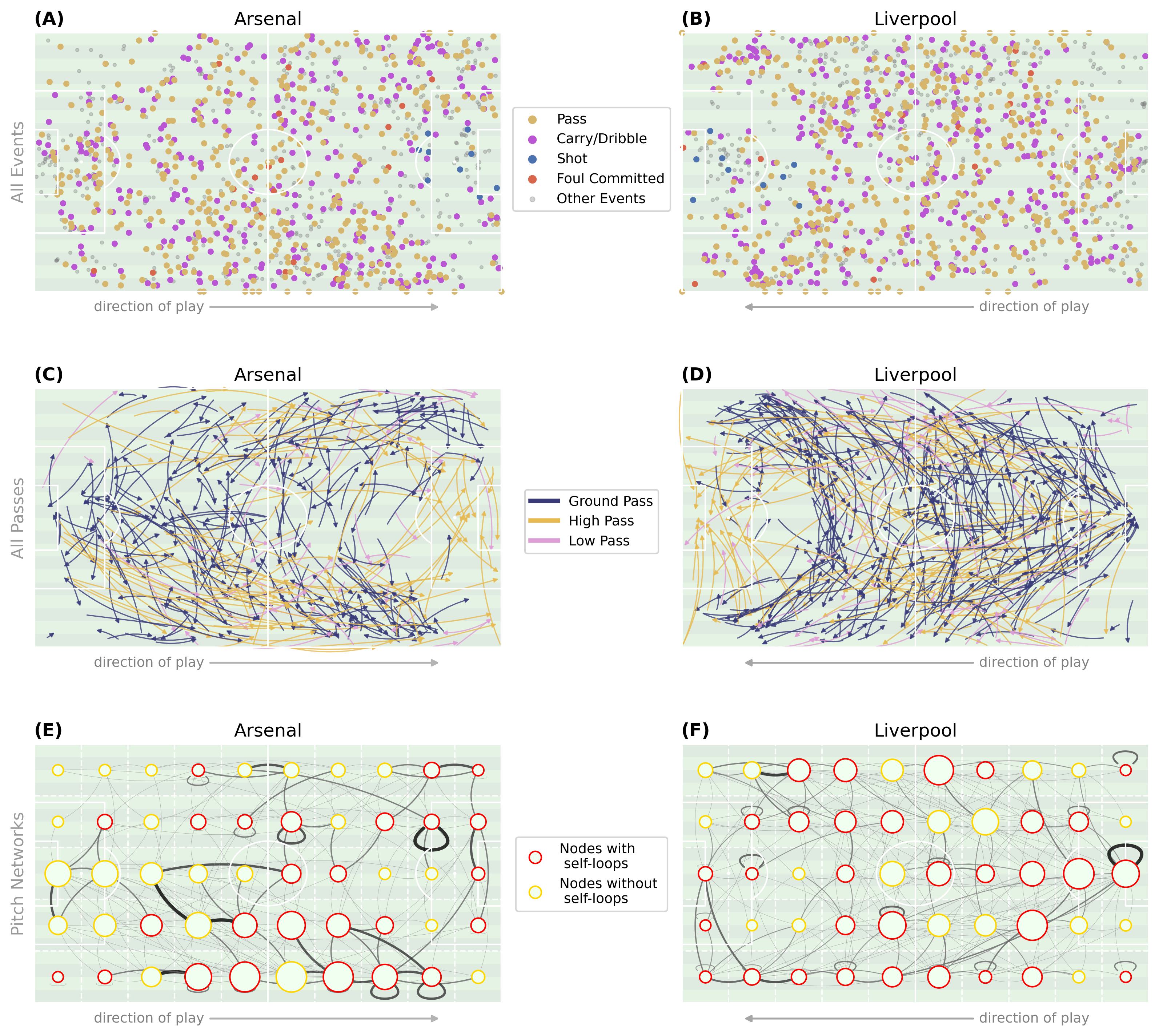}
\caption{\textbf{Pitch-passing network construction.} Schematic illustration of the transformation from raw match events to pitch-passing networks, shown for a representative match between Arsenal and Liverpool. \textbf{(A-B)} All recorded events including passes (beige), carries/dribbles (purple), shots (blue), fouls (red), and other actions (gray). \textbf{(C-D)} Spatial distribution of passes, color-coded by type: ground passes (blue), high passes (yellow), and low passes (pink). \textbf{(E-F)} Final pitch-passing networks constructed on a 10×5 grid (50 nodes). Node size reflects total passing activity; edge thickness indicates pass frequency between regions. Red nodes highlight regions with self-loops (multiple passes within the same zone); yellow nodes indicate regions without self-loops. The network representation captures spatial organization and directional flow of ball circulation.}
\label{fig:pitch_network}
\end{figure}

Our analysis integrates statistical performance indicators and network-based measures to characterize the evolution of elite football between 2020 and 2025. Figure~\ref{fig:pitch_network} illustrates our methodological approach, showing the transformation from raw match events to the pitch-passing networks used in our analysis. The results are organized into three parts: first, we examine passing and possession dynamics; next, we explore spatial organization and gameplay patterns; and finally, we investigate pitch-passing networks to uncover shifts in spatial control and organizational strategies.

\subsection*{Passing and possession dynamics}

\begin{figure}[t]
\centering
\includegraphics[width=\linewidth]{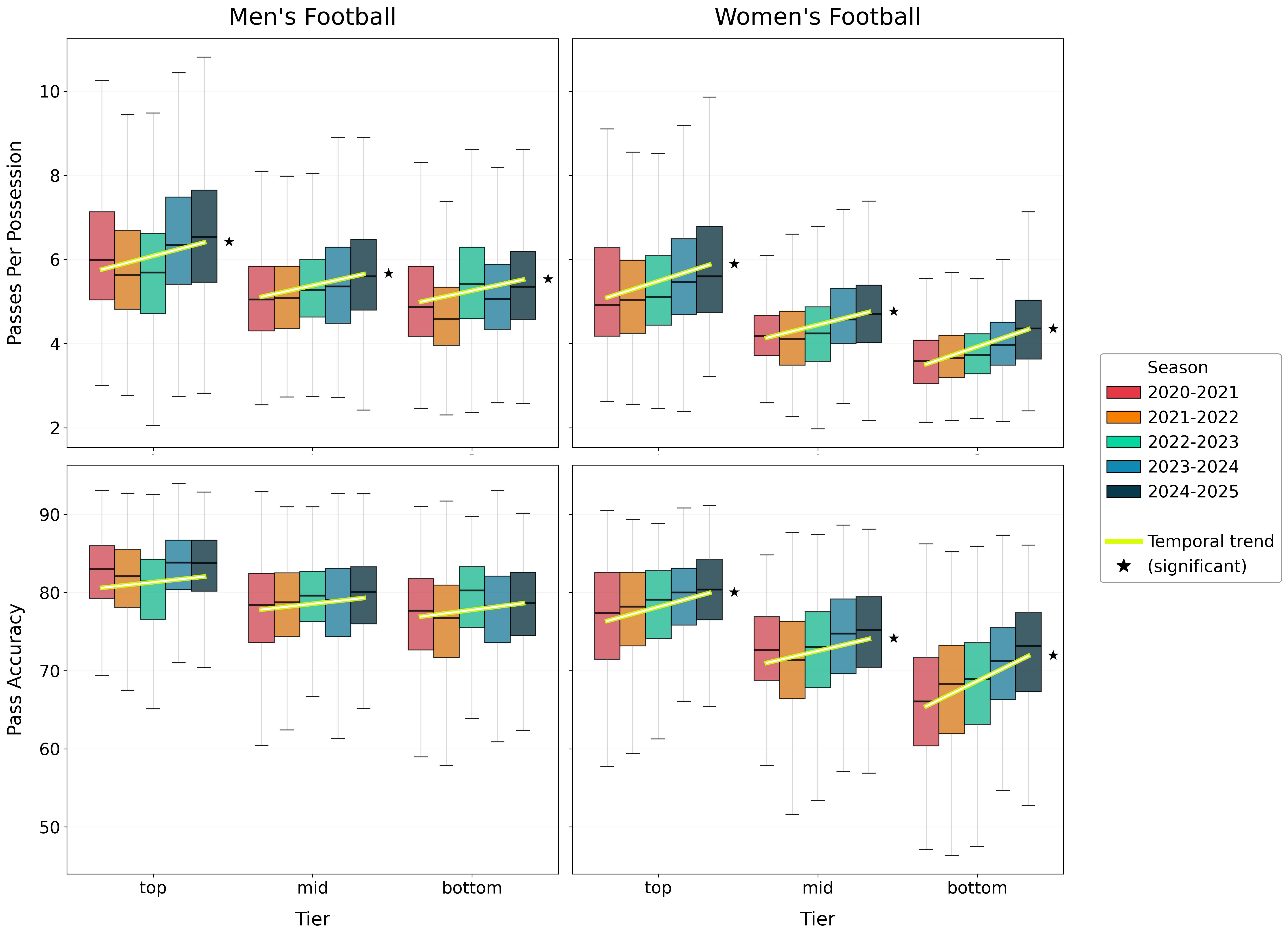}
\caption{\textbf{Evolution of possession intensity and passing precision.} Top: Passes per possession across team tiers for men's (left) and women's (right) football. Bottom: Overall pass accuracy across team tiers. Each panel shows distributions across five seasons (2020-2025), with temporal trends indicated by yellow lines. Asterisks ($\star$) denote statistically significant trends ($p < 0.05$ and absolute change > 5\%).}
\label{fig:ppp_accuracy}
\end{figure}

Figure~\ref{fig:ppp_accuracy} presents the evolution of two fundamental aspects of team possession: the intensity of ball circulation (passes per possession) and passing precision (overall accuracy). Passes per possession exhibits a clear performance hierarchy across all seasons, with top-tier teams averaging approximately 6.2 passes per possession in men's football and 5.5 in women's football, followed by mid-tier teams (5.4 and 4.5, respectively) and bottom-tier teams (5.2 and 3.9, respectively). Notably, all tiers in both men's and women's competitions show significant increasing trends over the five-year period, with women's football exhibiting particularly pronounced growth across all competitive levels (Figure~\ref{fig:ppp_accuracy}, top panels; Supplementary Table S4). This systematic increase suggests a league-wide shift toward more deliberate, possession-oriented play.

This pattern is also reflected in total passing volume (Supplementary Figure S2), where top teams complete approximately 540 passes per match compared to 390-450 for bottom-tier teams. Women's bottom-tier teams show a significant increasing trend in total passes, while other groups remain relatively stable. Additionally, passes before shot---the number of passes in sequences leading to goal attempts---exhibits consistent increases across all tiers in women's football and modest increases in men's football (Supplementary Figure S3), reinforcing the trend toward more elaborate attacking buildups.

Pass accuracy mirrors this performance hierarchy, with top teams achieving the highest completion rates (men: 81.9\%, women: 78.2\%), followed by mid-tier (men: 78.4\%, women: 72.6\%) and bottom teams (men: 77.4\%, women: 68.7\%) (Supplementary Table S5). Women's football demonstrates stronger temporal trends, with significant increases observed in top and bottom tiers (Figure~\ref{fig:ppp_accuracy}, bottom panels). When examining ground passes specifically---which constitute the majority of pass attempts---accuracy ranges from 90-95\% in men's leagues to 85-90\% in women's leagues, with both showing gradual improvements over time (Supplementary Figure S4). The gender gap in passing accuracy is consistent across tiers, amounting to roughly 4 percentage points in top teams and 9 percentage points in bottom teams (Supplementary Table S6), with larger relative gaps among lower-performing teams.

\begin{figure}[ht!]
\centering
\includegraphics[width=\linewidth]{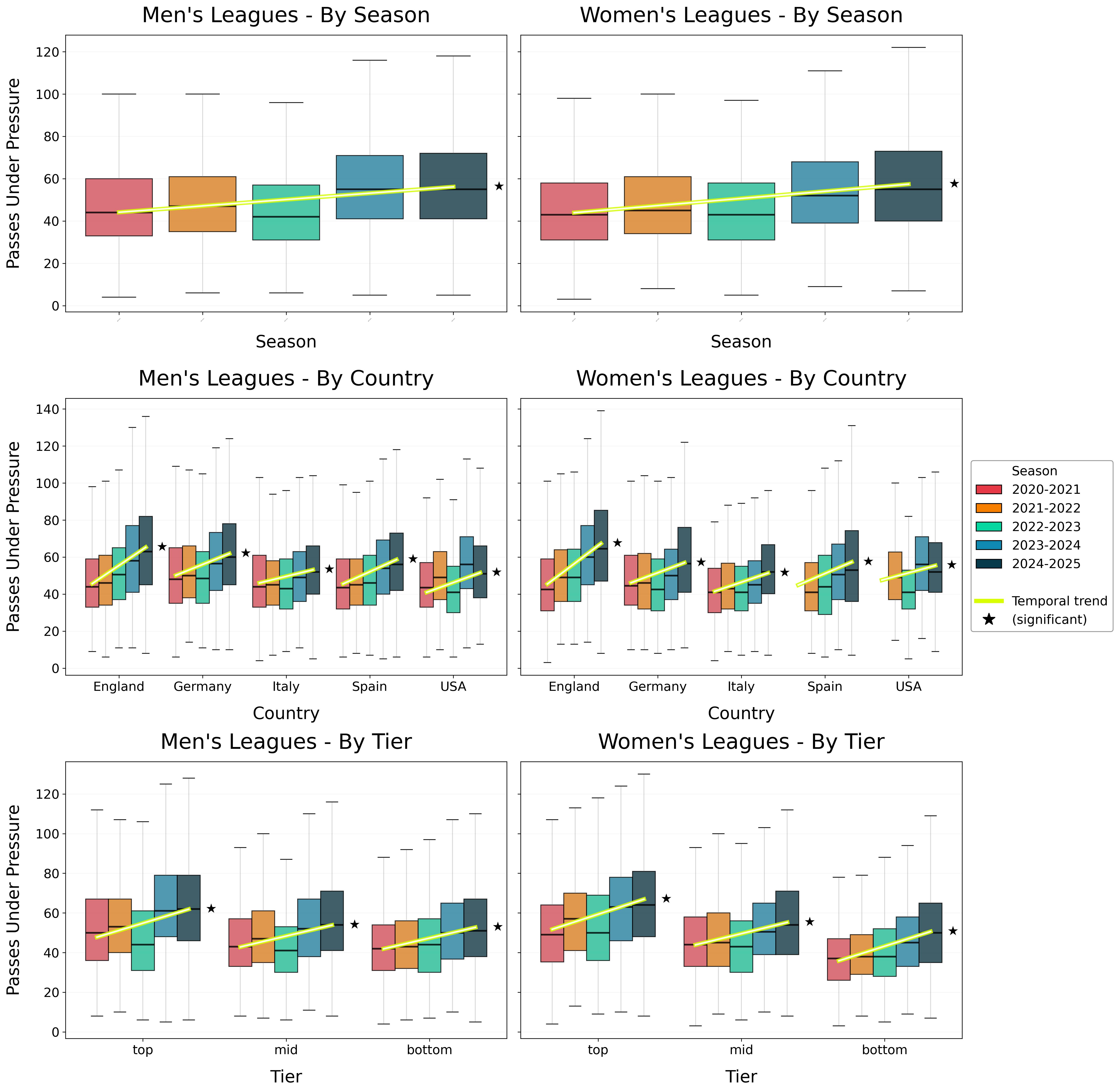}
\caption{\textbf{Intensification of play under defensive pressure.} Passes completed under pressure per match, shown for men's (left) and women's (right) football across all ten competitions. Each panel displays distributions split by season (top), by country (middle) and competitive tier (bottom). The colors indicate the season. Yellow trend lines with asterisks ($\star$) indicate significant temporal increases ($p < 0.05$ and change $> 5\%$). All the leagues show consistent upward trends across tiers, countries and genders.}
\label{fig:pressure}
\end{figure}

The intensification of play under defensive pressure represents another dimension of tactical evolution. Figure~\ref{fig:pressure} shows passes completed under pressure across all ten competitions, revealing substantial temporal increases in both men's and women's football. This trend is particularly pronounced and consistent in the English and Spanish competitions, where all tiers show significant year-over-year growth (Figure~\ref{fig:pressure}, top and middle panels). The magnitude of increase is similar across genders, with both men's and women's competitions experiencing approximately 3-4 additional passes under pressure per match annually across all tiers (Supplementary Table S4). Notably, top women's teams record higher absolute values than their male counterparts (60.3 vs. 55.0 passes under pressure per match; Supplementary Table S5), suggesting that the women's elite game already operates at greater pressure intensity at the highest competitive level.

Notably, league-specific patterns emerge in the data. German and Italian women's leagues exhibit stronger increases than their male counterparts, with the Frauen Bundesliga's mid-tier teams showing particularly dramatic growth in the 2024-2025 season (Figure~\ref{fig:pressure}, right panels). United States competitions follow a similar trend, consistent with a broad, sport-wide intensification of play under pressure.

\subsection*{Spatial organization and gameplay patterns}

The possession dynamics described above capture \textit{how much} and \textit{how accurately} teams pass, but not \textit{where} on the pitch play unfolds. To address this, we examined three complementary aspects of spatial organization and match flow: offside positioning, directional play, and spatial pass positioning.

\begin{figure}[t]
\centering
\includegraphics[width=\linewidth]{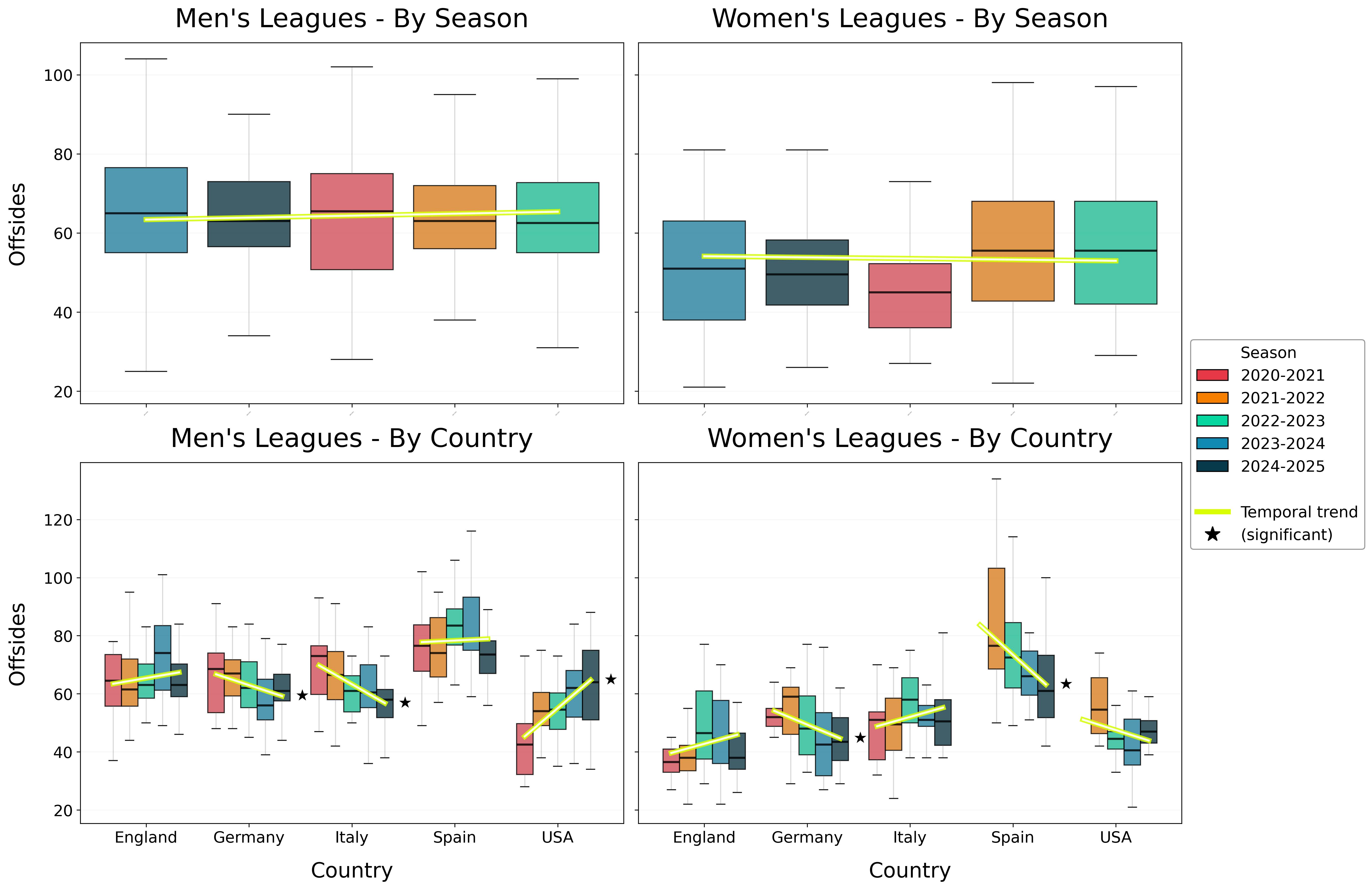}
\caption{\textbf{Spatial organization and gameplay dynamics.} Offside infractions accumulated per team over the course of each season (i.e., summed across all matches), shown by season (top) and country (bottom). All panels compare men's (left) and women's (right) football across five seasons. Yellow trend lines with asterisks indicate significant changes ($p < 0.05$ and change > 5\%). Note the pronounced country-level differences in offsides, with Spain exhibiting substantially higher rates than other nations.}
\label{fig:gameplay}
\end{figure}

Offside patterns exhibit clear country-level differences but minimal temporal variation (Figure~\ref{fig:gameplay}; Supplementary Table S7). Spanish leagues record higher offside rates than the other countries pooled in both men's (2.06 vs 1.71, $p < 10^{-4}$) and women's football (2.54 vs 2.10, $p < 10^{-4}$), representing an approximately 20\% increase. This persistent difference is consistent with stylistic differences that have been widely discussed for Spanish football (although we do not directly measure defensive line height in this dataset).

\begin{figure}[t]
\centering
\includegraphics[width=\linewidth]{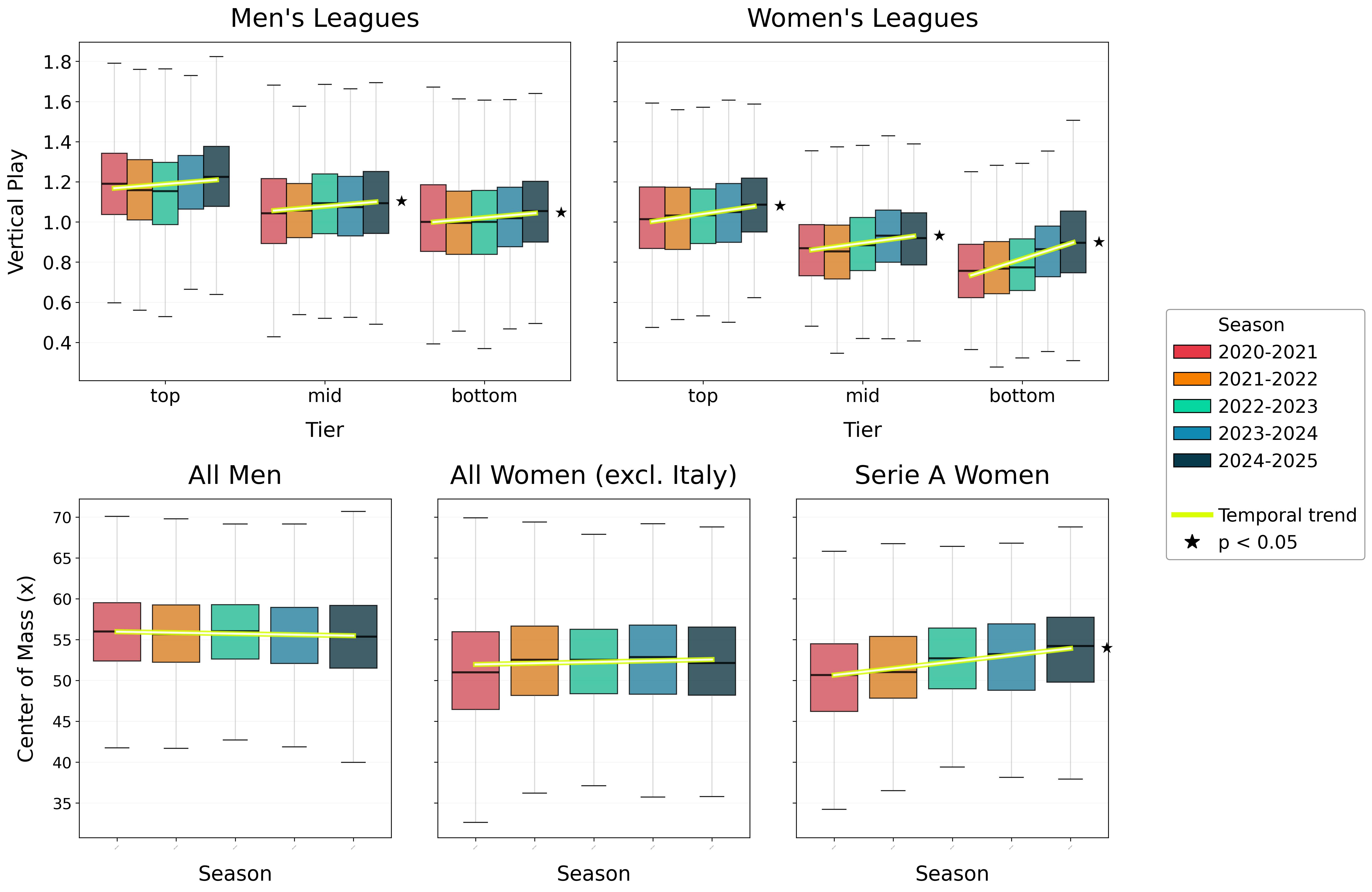}
\caption{\textbf{Evolution of spatial positioning: Vertical play and pass center of mass.} Top row: ratio of forward-directed to lateral passes across the pitch by tier, with seasons indicated by color. Bottom row: longitudinal (x-coordinate) position of pass origins for three groups: all men's teams (left), all women's teams excluding Italy (center), and Serie A Women (right). A notable increase in the percentage of vertical passes is observed across seasons, competitions, and tiers. While men's football shows declining attacking depth and other women's leagues show minimal change, Serie A Women exhibits a strong progressive increase in forward positioning, suggesting rapid tactical modernization within this competition.}
\label{fig:forward}
\end{figure}

Vertical play, defined as the ratio of forward-directed to lateral passes, reveals systematic differences across competitive levels (Figure~\ref{fig:forward}, top row; Supplementary Table S5). Top-tier teams exhibit the highest proportion of vertical passes in both men's (ratio = 1.2, i.e., 20\% more forward than lateral passes) and women's (1.0) football, followed by mid-tier and bottom teams. This hierarchy suggests that stronger teams employ more direct, goal-oriented passing patterns. Temporal trends show significant increases in women's football, particularly among top and bottom-tier teams (Supplementary Table S4), indicating an ongoing shift toward more vertical gameplay in women's competitions. Men's football shows more stable patterns with only modest changes over time.

Throw-in length shows modest tier-based differences and divergent temporal trends between genders, with men's leagues exhibiting stable or increasing distances and women's leagues displaying consistently decreasing distances (see Supplementary Figure S6). Shot distance exhibits a consistent declining trend across most tiers in men's football, with top and mid-tier teams showing significant decreases of approximately 0.14-0.15m per year (Supplementary Figure S5; Supplementary Table S4). Women's football shows similar declining patterns, though not reaching statistical significance. These trends suggest teams are working the ball closer to goal before shooting, aligning with the observed increases in passes before shot.

Spatial positioning patterns reveal distinct evolutionary trajectories across competitions. Figure~\ref{fig:forward} presents the longitudinal center of mass---the average horizontal position of successful passes---across three groups: all men's teams, all women's teams excluding Italy, and Serie A Women separately. Men's football shows a modest but significant decrease in attacking depth over time (slope = -0.12 m/year, $p < 10^{-4}$), suggesting teams are initiating passes slightly deeper in their own half. Women's football excluding Italy exhibits a small positive trend (slope = +0.15 m/year, $p = 0.03$), indicating gradual forward progression (see Supplementary Table S8).

In contrast, Italy's Serie A Women shows a clear temporal increase (slope = +0.82 units/year, $p < 10^{-4}$), with the median longitudinal position advancing by approximately 4 units over the five-year period. This pronounced forward progression --- distinct from the modest or stable trends observed in all other groups --- signals rapid tactical modernization within this competition, likely reflecting a convergence of investment growth, evolving coaching philosophies, and institutional restructuring that warrants dedicated future investigation. More broadly, this league-specific trajectory underscores that women's football does not follow a uniform developmental path: tactical sophistication emerges through diverse, locally shaped processes, making it essential to examine individual competitions rather than treating women's leagues as a monolithic category.

\subsection*{Pitch-passing network analysis}

\begin{figure}[t]
\centering
\includegraphics[width=\linewidth]{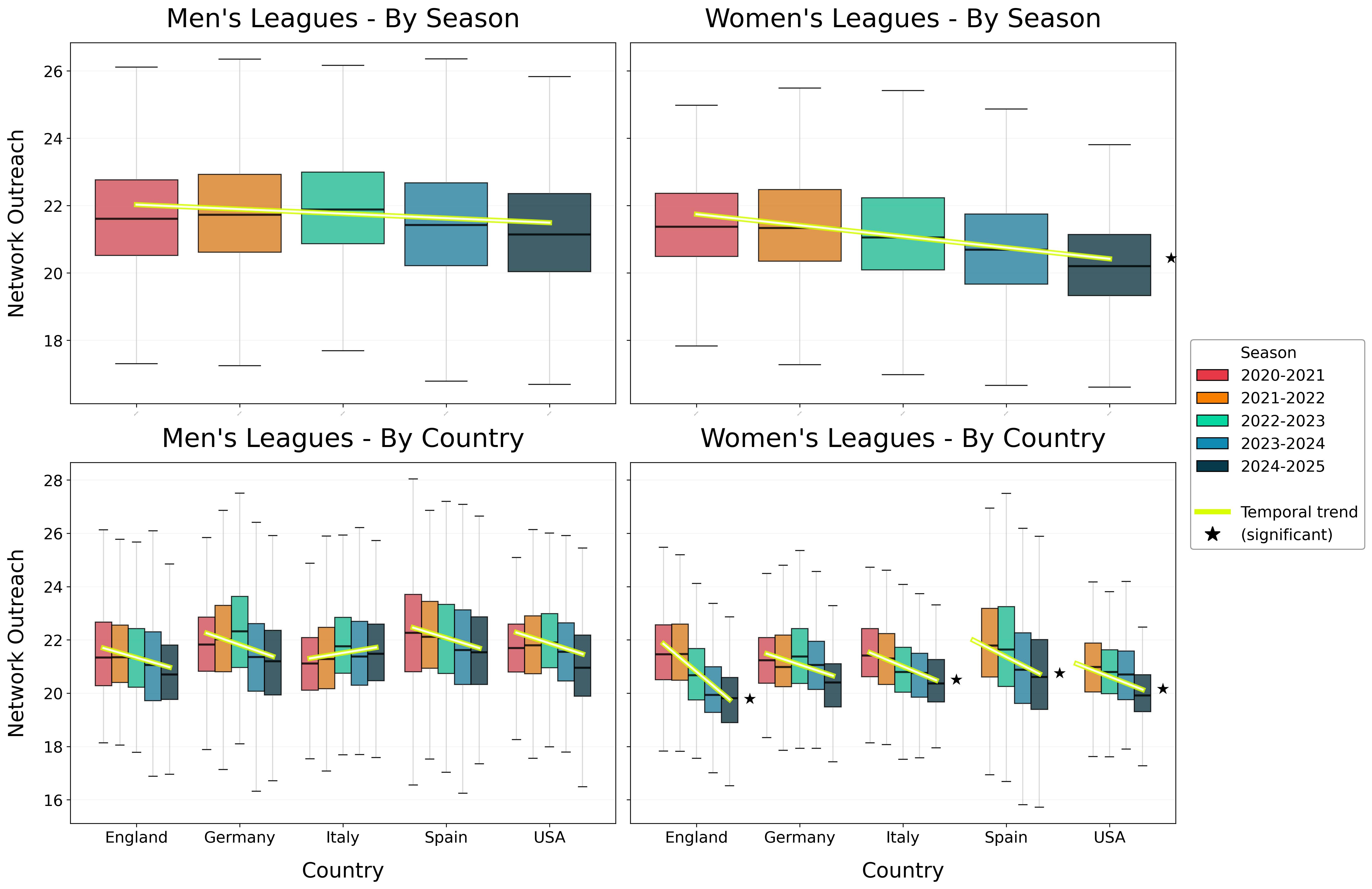}
\caption{\textbf{Spatial organization of ball circulation.} Normalized network outreach across seasons (top panels) and countries (bottom panels) for men's (left) and women's (right) football. Lower outreach indicates more concentrated, shorter-range passing connections. Yellow trend lines with asterisks mark significant declining trends ($p < 0.05$ and change > 5\%), most pronounced in women's competitions. Country-level differences reveal persistent national tactical signatures, with Spain maintaining the highest spatial dispersion.}
\label{fig:network_outreach}
\end{figure}

To capture the spatial organization of ball circulation, we constructed pitch-passing networks for each team and computed structural measures characterizing connectivity patterns. Figure~\ref{fig:network_outreach} presents normalized network outreach---a measure combining passing frequency and spatial distance that quantifies how widely the ball circulates across the pitch---across seasons and countries.

Network outreach exhibits a consistent declining trend across both genders and all countries over the five-year period (Figure~\ref{fig:network_outreach}, top panels; Supplementary Table S4). This decrease is particularly pronounced in women's football, where all countries show significant negative trends. The temporal pattern indicates that teams are increasingly concentrating ball circulation into shorter-range passing connections rather than distributing play widely across distant pitch regions. This shift aligns with the observed decrease in average pass length and increase in passes per possession reported earlier, collectively suggesting a tactical evolution toward shorter, more intricate passing sequences.

These spatial changes are accompanied by shifts in network structural properties. Maximum eigenvalue of the adjacency matrix---reflecting connectivity strength and hierarchical organization---shows a clear tier-based hierarchy (top > mid > bottom) in both genders (Supplementary Table S5), with women's bottom-tier teams exhibiting significant increases over time (Supplementary Table S4). Average shortest path length, quantifying circulation efficiency, remains relatively stable across most groups (Supplementary Figure S7), indicating that despite more concentrated spatial distribution, teams maintain efficient connectivity across the pitch. Top teams consistently achieve the shortest path lengths (approximately 3.2-3.4), enabling rapid ball progression to any region.

Country-level patterns reveal systematic differences in spatial organization (Figure~\ref{fig:network_outreach}, bottom panels). Spanish leagues maintain the highest network outreach in both genders, consistent with Spain's traditional emphasis on expansive, possession-based football. In contrast, English and Italian leagues exhibit lower outreach values, reflecting more compact passing structures. These national differences remain stable across the five-year period despite the overall declining trend, indicating that while absolute outreach values decrease, relative differences between countries' tactical styles persist.

\begin{figure}[t]
\centering
\includegraphics[width=0.8\linewidth]{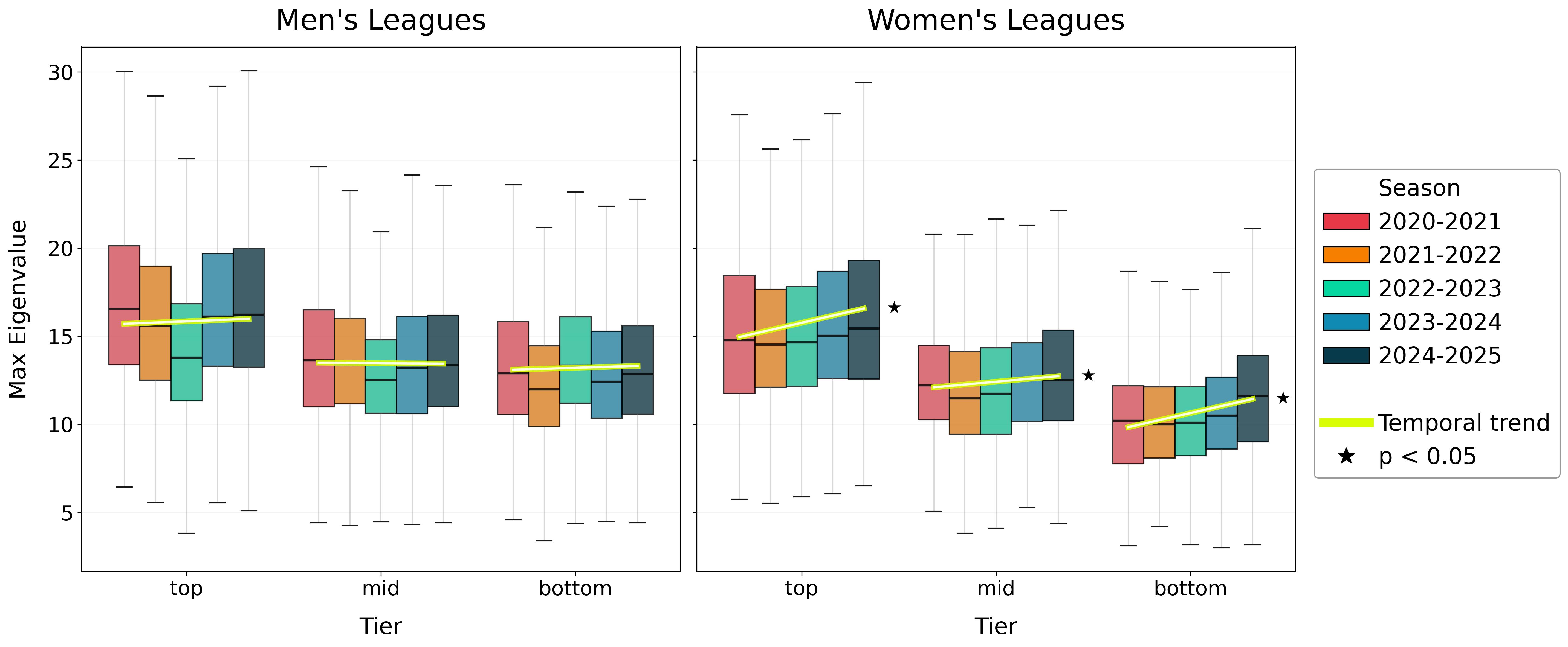}
\caption{\textbf{Passing network structure and hierarchical organization.} Maximum eigenvalue of pitch-passing network adjacency matrix by tier for men's (left) and women's (right) football. Higher values indicate stronger, more centralized network structures with dominant passing routes acting as hub connections. The consistent tier gap across both genders reflects a structural signature of competitive level. Women's bottom-tier teams exhibit a significant increasing trend, suggesting growing network cohesion at lower competitive levels.}
\label{fig:eigenvalue}
\end{figure}

When disaggregated by tier, an inverse relationship emerges between team performance and network outreach (Supplementary Table S5): top teams exhibit the lowest outreach (men: 21.2, women: 20.4), followed by mid-tier (men: 22.0, women: 21.2) and bottom teams (men: 22.1, women: 21.4). This inverse relationship indicates that elite performance is associated with spatially compact ball circulation rather than broad territorial coverage. Together with their higher eigenvalues and shorter path lengths (Figure \ref{fig:eigenvalue} and Supplementary Figure S7), these results paint a picture of top teams constructing more centralized, structurally cohesive passing networks that enable rapid ball progression through compact spatial organization.

Maximum eigenvalue of the network adjacency matrix reflects overall connectivity strength and hierarchical organization (Figure~\ref{fig:eigenvalue}). Top-tier teams exhibit substantially higher eigenvalues (men: 16.4, women: 15.8) than mid-tier (13.6, 12.5) and bottom-tier teams (13.1, 10.7), indicating more robust and centralized passing structures. The pronounced gap between tiers points to a structural signature of competitive level, with stronger teams organizing ball circulation around a more dominant network backbone. Women's bottom-tier teams show a significant increasing trend, suggesting improving network cohesion at lower competitive levels.

\section*{Discussion}

Our analysis of 13,018 matches across ten elite competitions shows a systematic evolution in football gameplay between 2020 and 2025, characterized by four interconnected trends that collectively reshape how the sport is played at the highest level. First, passes completed under defensive pressure have increased substantially across all tiers and genders, indicating a sport-wide intensification toward more physically demanding play. This trend reflects not merely increased fitness levels but a fundamental tactical shift in which teams actively seek to constrain opponents' decision-making time and space. Second, teams are working the ball progressively closer to goal before shooting, suggesting more patient, controlled approaches to chance creation rather than speculative long-range attempts. Third, passes per possession continue their documented rise \cite{gonzalezrodenas2023tactics}, extending a trend observable since 2008 and indicating sustained evolution toward more elaborate buildup sequences. Fourth, network outreach has declined consistently, particularly in women's football, reflecting more concentrated spatial ball circulation. Taken together, these patterns point to a fundamental strategic paradigm: modern football increasingly prioritizes sustained possession through shorter, more intricate passing sequences, coupled with intensified pressing when out of possession---tactical principles exemplified by contemporary coaching philosophies \cite{violan2014}.

The observed improvements in passing accuracy --- particularly in ground passes, which constitute the majority of attempts --- likely reflect multiple interacting factors, including tactical preferences (e.g., shorter passing sequences), technical training, and coaching emphasis. Because we do not directly measure pitch conditions, training environments, or rule and officiating changes over time, we avoid attributing these trends to any single mechanism. That said, pitch quality improvements over the study period likely contribute: modern elite stadiums employ sophisticated drainage systems, hybrid grass-synthetic surfaces, and intensive maintenance regimes that maintain consistent playing surfaces, facilitating the execution of technical skills and supporting the precision-based passing game that characterizes contemporary football. The gender gap in passing accuracy amounts to roughly 4 percentage points in top-tier teams and 9 percentage points in bottom-tier teams (Supplementary Table S6), which may reflect not only differences in technical development pathways but also potential disparities in training facility quality and pitch maintenance standards between men's and women's competitions, though the latter is rapidly improving as women's football professionalization accelerates.

Women's football exhibits particularly pronounced rates of change across multiple metrics, consistent with a competition undergoing rapid professionalization and tactical development \cite{fifa2023}. The magnitude of temporal trends in women's competitions often exceeds those observed in men's football, suggesting that the women's game is experiencing accelerated tactical evolution as leagues mature, coaching methodologies become more sophisticated, and player development pathways professionalize. Critically, our analysis reveals substantial heterogeneity in these evolutionary trajectories. Italy's Serie A Women demonstrates a pronounced forward progression in pass positioning (Figure~\ref{fig:forward}; Supplementary Table S8)---advancing approximately four meters upfield over five seasons---distinct from patterns observed in other women's competitions. This league-specific modernization underscores that women's football does not follow a uniform developmental path but rather exhibits distinct trajectories shaped by local investment patterns, coaching philosophies, and institutional structures. Beyond its descriptive value, this heterogeneity carries methodological implications: aggregating women's leagues into a single category risks masking divergent evolutionary dynamics and may lead to misleading cross-gender comparisons. As women's football continues to professionalize at different rates across countries, future research and policy decisions --- including resource allocation, coaching certification standards, and league governance --- would benefit from frameworks that treat individual competitions as distinct systems rather than instances of a common template.

Network analysis reveals that elite performance emerges from qualitatively different spatial organization rather than mere quantitative superiority. Top teams construct more centralized, efficient passing networks---evidenced by higher eigenvalues and shorter path lengths---yet paradoxically exhibit lower network outreach than bottom-tier teams. This counterintuitive finding suggests that success derives from compact, cohesive ball circulation that enables rapid progression through concentrated spatial zones, rather than wide distribution across the pitch. The declining outreach observed across all competitions indicates a sport-wide tactical convergence: teams increasingly prioritize controlled possession in concentrated areas over expansive territorial coverage, aligning with the broader trend toward shorter passes and extended possession sequences.

Country-level differences reveal that while tactical evolution affects all competitions, national styles remain distinguishable throughout. Spain's persistently higher offside rates and network outreach across both genders reflect characteristic high defensive lines and expansive possession play. These stable national signatures, maintained despite overall league-wide trends, suggest that football's tactical evolution operates within culturally constrained boundaries---leagues modernize and intensify, yet distinctive national identities persist.

Several limitations warrant acknowledgment and possibly further research. Our five-season analytical window, while capturing recent evolution, may not reveal longer-cycle tactical shifts that characterize multi-decade developments in established men's leagues. The proprietary nature of event data, though necessary for comprehensive analysis, may limit reproducibility. Future research should extend the temporal scope to contextualize recent changes within longer historical trajectories, normalize metrics by effective playing time, which could meaningfully affect comparisons given known variations in ball-in-play duration across leagues \cite{spyrou2023,altmann2023,tojo2023}, investigate team-specific tactical signatures that may be obscured in league-level aggregation using advanced network motif analysis \cite{li2025,aguirre2013}, and explore whether observed trends predict competitive success or represent stylistic preferences independent of outcomes \cite{basini2023}. The rapid transformation of women's football---marked by accelerating professionalization, increasing investment, and evolving tactical sophistication---merits dedicated longitudinal study to document this historic developmental phase. Additionally, institutional structures such as ownership models \cite{budzinski2020} and wearable technology adoption \cite{seckin2023} may influence tactical evolution in ways not captured by our analysis.

Overall, our results provide quantitative evidence of recent shifts in possession, pressure, and spatial passing structure across multiple elite competitions. The consistency of several trends across countries and tiers suggests that recent tactical changes are not confined to a single league, while persistent cross-country differences indicate that stylistic diversity remains.

\section*{Methods}

\subsection*{StatsBomb Events Data}

The data analyzed in this work, provided under license from Hudl StatsBomb \cite{statsbomb_private_2025}, comprise 13,018 matches played between 2020 and 2025 across ten top-tier competitions in five countries (Supplementary Table S1). The dataset consists of event data---timestamped records of all on-pitch actions including their Euclidean coordinates and outcomes---covering passes, shots, fouls, substitutions, and other match events. Our analysis focuses primarily on passing and shooting actions, as these capture the fundamental dynamics of ball movement and goal-scoring opportunities. 

We classified teams into three performance tiers (top, mid, bottom) based on percentile ranks of final league standings within each season and competition, dividing teams into approximately equal thirds. This classification enables comparative analysis of tactical evolution across different competitive levels while accounting for varying league sizes.

\subsection*{Statistical analysis}

We computed a comprehensive set of team-level performance metrics for each match, encompassing passing dynamics, shooting behavior, spatial positioning, and game flow characteristics. From this initial set, we selected metrics that exhibited meaningful variation, were non-redundant with one another, and provided insight into the tactical and strategic evolution of football. A complete list of all computed statistical features is provided in Supplementary Table S2.

Additionally, we derived several spatial and tactical features: (i) \textit{pass center of mass} (mean $x$-coordinate of completed pass origins), capturing average attacking depth; (ii) \textit{vertical play}, defined as the ratio of forward-directed to lateral passes, where a pass is classified as forward if its displacement along the pitch axis (x-component) exceeds its lateral displacement (y-component), and as lateral otherwise; and (iii) \textit{offsides} (number of offside infractions per team per season).

Our statistical analysis addressed three complementary questions. First, to assess {temporal evolution within tiers}, we performed linear regression for each combination of gender and tier (top, mid, bottom), using season year as the independent variable and the metric of interest as the dependent variable. Regression was performed on all individual team-season observations rather than aggregated means, preserving the full distribution of the data. This approach quantifies whether performance indicators changed systematically over the five-year period (2020-2025) and whether the rate of change differed across competitive levels. In visualizations, temporal trends are marked as statistically significant (indicated by $\star$) only when both $p < 0.05$ and the absolute percent change over the five-year period exceeded 5\%.

Second, to evaluate {differences among tiers}, we conducted one-way ANOVA separately for men's and women's competitions, testing whether mean metric values differed significantly across the three tiers within each gender. Third, to compare {gender differences by tier}, we performed independent-samples t-tests for each tier, comparing men's and women's teams at the same competitive level. For each comparison, we report the mean values for both groups, the absolute difference, and the percent difference calculated as $\Delta\% = 100 \times (M - W) / W$, where positive values indicate higher performance in men's football.

All statistical tests were two-tailed with significance threshold set at $p < 0.05$. Results are summarized in Tables S4, S5, and S6 with detailed outputs provided in the Supplementary Information.

\subsection*{Pitch network}

\subsubsection*{Network construction}

To complement statistical performance indicators, we constructed pitch-passing networks for each team in every match. This network-based representation has been successfully applied in previous studies to identify playing styles, quantify spatial organization, and reveal tactical patterns not readily apparent from conventional statistics \cite{garrido2020, zhou2023, herrera2020, novillo2024, huang2024}.

The pitch was discretized into a regular grid of spatial zones using StatsBomb's standardized $120 \times 80$ coordinate system. Following established methodologies \cite{garrido2020}, we employed a $10 \times 5$ grid structure, yielding 50 spatial regions (Supplementary Figure S1). This configuration aligns with prior work demonstrating that approximately 49 regions represent an optimal balance between spatial resolution and statistical reliability for pitch network analysis \cite{garrido2020}. Each region serves as a node in the network, while directed, weighted edges connect regions between which successful passes occur. Edge weights correspond to the number of passes exchanged between regions throughout the match, capturing both the frequency and directionality of ball movement across the pitch.

To facilitate comparative analysis across matches with different passing volumes, we normalized edge weights by the total network strength (sum of all edge weights), yielding a representation in which link weights sum to 100. This normalization ensures that network metrics reflect structural properties of ball circulation rather than overall passing volume. Figure~\ref{fig:pitch_network} illustrates this construction process for a representative match between Arsenal and Liverpool.

\subsubsection*{Network analysis}

We computed network-level metrics that characterize the structural properties of ball circulation patterns. \textit{Average shortest path length} quantifies the average minimum number of passes required to connect any two regions, with lower values indicating more direct ball progression. \textit{Maximum eigenvalue} of the adjacency matrix reflects the network's overall connectivity and hierarchical structure, with higher values suggesting more centralized passing patterns. \textit{Outreach} measures the spatially weighted dispersion of passes, calculated as the sum of pass counts multiplied by the Euclidean distance between connected regions, normalized by total network strength; higher outreach indicates longer-range ball circulation across the pitch.

These network metrics were aggregated at the team-season level and analyzed across tiers, genders, and countries following the same statistical framework described above. Complete descriptions of all network measures are provided in Supplementary Table S3.

\section*{Acknowledgements}

This work is the output of the Complexity72h workshop, held at University Carlos III of Madrid, Spain, 23-27 June 2025; \url{https://www.complexity72h.com}. J.M.B. and P. R.-S. acknowledge the support by Ministerio de Ciencia e Innovación, Spain under grant PID2023-147827NB-I00.

\section*{Author contributions statement}

MT, BK, and JMB conceived the study. RC, RD, PE, NF, SLN, RO, LKR, PRS, LS, and MT conducted the experiments and data analysis. All authors analyzed the results and contributed to manuscript writing. MT, BK, and JMB reviewed and edited the final manuscript. All authors reviewed and approved the manuscript.

The authors declare no competing interests.

\section*{Data availability}

The football match event data analyzed in this study were obtained from StatsBomb Services Ltd under institutional license and are not publicly available. Python code for the statistical and network analyses is publicly available at \url{https://github.com/maddaleona/harder_shorter_sharper_forward/}.

\bibliography{sample}


\newpage

\begin{center}
{\sffamily\bfseries\fontsize{20}{25}\selectfont
Supplementary Information for:\\
Harder, shorter, sharper, forward: A comparison of women's and men's elite football gameplay (2020-2025)}

\vspace{10pt}

{\sffamily\fontsize{12}{16}\selectfont
Rebecca Carstens$^1$, Raj Deshpande$^2$, Pau Esteve$^3$, Nicolò Fidelibus$^4$, Sara Linde Neven$^5$, Ramona Ottow$^6$, Lokamruth K. R.$^7$, Paula Rodríguez-Sánchez$^8$, Luca Santagata$^{9,10}$, Javier M. Buldú$^8$, Brennan Klein$^{10,11,12}$, Maddalena Torricelli$^{10,*}$}
\end{center}

\vspace{25pt}

\section*{Dataset composition}

Table~\ref{tab:SI_dataset} provides a detailed breakdown of the 13,018 matches analyzed in this study, organized by country and gender. The dataset spans five seasons (2020-2025) and encompasses both established leagues with stable team numbers (European competitions) and expanding leagues (MLS and NWSL in the United States).

\begin{table}[ht]
\centering
\begin{tabular}{|l|l|cc|c|}
\hline
\textbf{Country} & \textbf{Competition} & \textbf{Men} & \textbf{Women} & \textbf{Total} \\
\hline
Germany & Bundesliga / Frauen Bundesliga & 1,529 & 652 & 2,181 \\
Spain & La Liga / Liga F & 1,900 & 927 & 2,827 \\
England & Premier League / FA WSL & 1,899 & 659 & 2,558 \\
Italy & Serie A / Serie A Women & 1,900 & 620 & 2,520 \\
USA & MLS / NWSL & 2,276 & 656 & 2,932 \\
\hline
\multicolumn{2}{|l|}{\textbf{Total}} & \textbf{9,504} & \textbf{3,514} & \textbf{13,018} \\
\hline
\end{tabular}
\caption{\textbf{Supplementary Table S1.} Number of matches analyzed by country and gender (2020-2025). MLS expanded from 26 to 30 teams during this period; NWSL expanded from 10 to 14 teams. Liga F coverage began in the 2021-22 season.}
\label{tab:SI_dataset}
\end{table}

\begin{figure}[t]
\centering
\includegraphics[width=0.9\linewidth]{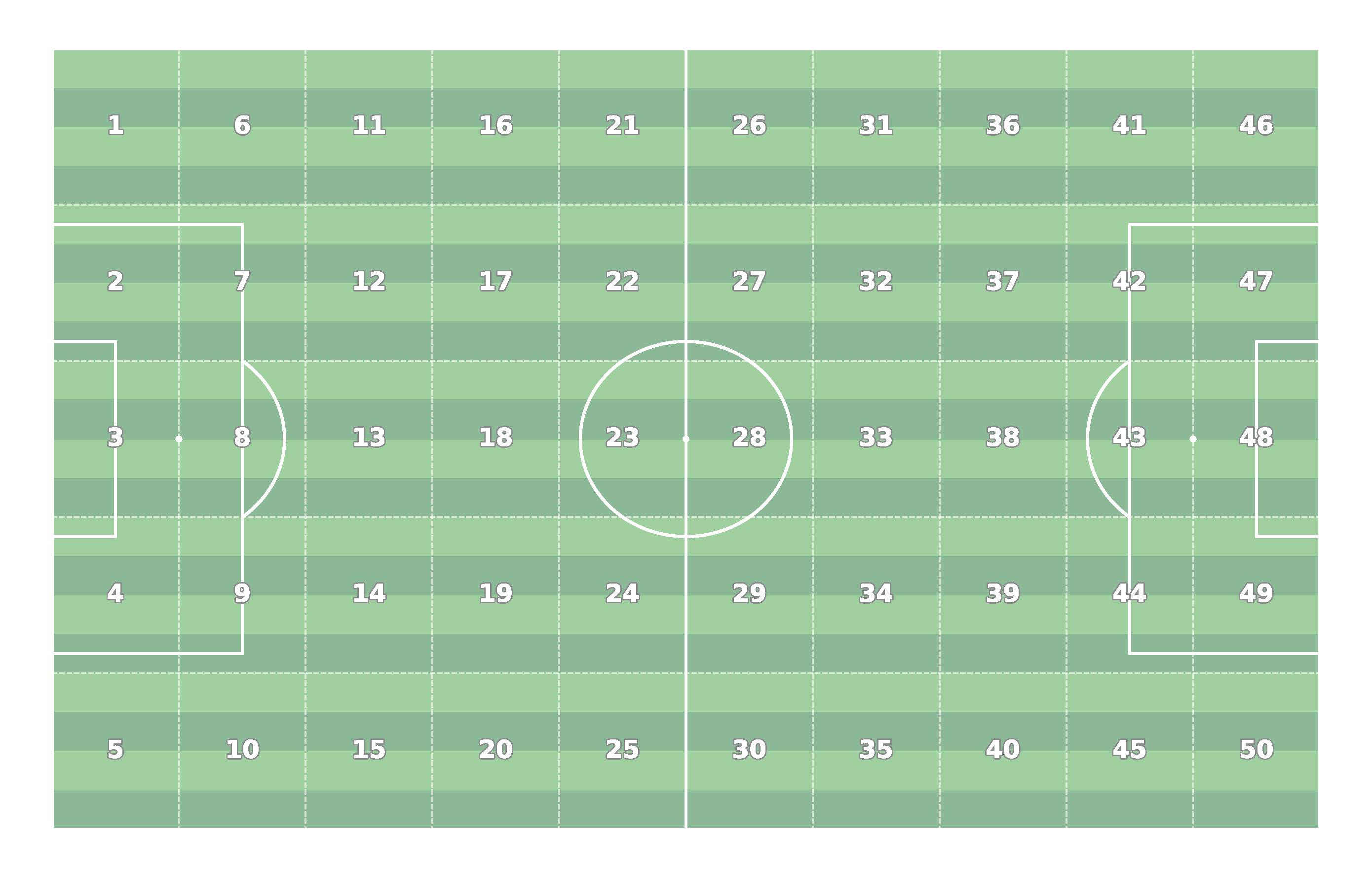}
\caption{\textbf{Supplementary Figure S1.} Spatial discretization of the football pitch for network construction. The pitch is divided into a 10×5 regular grid creating 50 zones numbered sequentially from the defensive third (left, nodes 1-10) to the attacking third (right, nodes 41-50). Each zone serves as a node in the pitch-passing network, with directed edges connecting zones between which successful passes occur. Edge weights correspond to the number of passes exchanged between regions throughout the match.}
\label{fig:SI_pitch_grid}
\end{figure}

\section*{Pitch network construction}

The construction of pitch-passing networks follows a spatial discretization approach in which the football pitch is divided into regular zones that serve as network nodes. Figure~\ref{fig:SI_pitch_grid} illustrates the 10×5 grid structure employed in this study, yielding 50 spatial regions across the standardized 120m × 80m pitch dimensions. Nodes are numbered sequentially from left to right (attacking direction), with rows representing different lateral positions on the pitch. Each successful pass between regions creates a directed edge in the network, with edge weight corresponding to the number of passes exchanged between those regions throughout the match.

\section*{Performance metrics}

We computed a comprehensive set of performance indicators from match event data, encompassing both conventional match statistics and network-derived measures of spatial ball circulation. Tables~\ref{tab:SI_stat_metrics} and~\ref{tab:SI_network_metrics} provide complete specifications for statistical and network-based measures, respectively, distinguishing between metrics reported in the main analysis and those computed for exploratory purposes.

\subsection*{Statistical metrics}

In Table \ref{tab:SI_stat_metrics}, we list the statistical measures analyzed in this work.

\begin{table}[ht]
\centering
\small
\begin{tabular}{|p{4.5cm}|p{8cm}|p{2cm}|}
\hline
\textbf{Metric} & \textbf{Description} & \textbf{Unit} \\
\hline
\multicolumn{3}{|c|}{\textit{Reported in article}} \\
\hline
Number of passes & Total number of passes attempted by the team during the match. & count \\
\hline
Passes per possession (PPP) & Average number of completed passes during each team possession phase; indicates possession style (buildup vs. direct play). & passes \\
\hline
Pass accuracy & Percentage of attempted passes that were successfully completed. & \% \\
\hline
Ground pass accuracy & Percentage of ground-level passes that were successfully completed. & \% \\
\hline
Passes before shot (PBS) & Average number of passes completed by the team in the sequence leading to each shot attempt. & passes \\
\hline
Passes under pressure & Number of completed passes made while the passer was under defensive pressure. & count \\
\hline
Shot distance & Average Euclidean distance from shot initiation location to the center of the goal. & meters \\
\hline
Vertical play & Ratio of vertical passes (toward opponent's goal) to horizontal passes (lateral movement). & ratio \\
\hline
Throw-in length & Average distance between throw-in location and receiving player position. & meters \\
\hline
Offsides & Total number of offside infractions committed by the team. & count \\
\hline
Pass center of mass (x) & Mean horizontal (longitudinal) coordinate of all successful pass starting locations; proxy for average attacking depth. & meters \\
\hline
\multicolumn{3}{|c|}{\textit{Additional metrics computed}} \\
\hline
Total events & Total number of all recorded events (passes, shots, fouls, etc.) by the team. & count \\
\hline
Number of shots & Total number of shots taken by the team. & count \\
\hline
Spatial entropy & Normalized average distance between consecutive successful pass locations. & --- \\
\hline
High/Low pass accuracy & Accuracy of aerial and bouncing passes, respectively. & \% \\
\hline
Max/Avg pass length & Maximum and average length of completed passes. & meters \\
\hline
Number of substitutions & Number of player substitutions made by the team. & count \\
\hline
Shot accuracy & Percentage of shots on target (blocked, saved, or goal). & \% \\
\hline
Average shot xG & Average expected goals value per shot attempt. & xG \\
\hline
Shot execution xG uplift & Average difference between post-shot xG and pre-shot xG. & xG \\
\hline
Average shot angle & Average angle subtended by the two goalposts from shot location. & degrees \\
\hline
Pass center of mass (y) & Mean lateral (transverse) coordinate of successful pass starting locations. & meters \\
\hline
Possession & Percentage of match time during which the team controlled the ball. & \% \\
\hline
Max PPP / Max PBS & Maximum passes per possession / before shot in any single sequence. & passes \\
\hline
Effective time & Duration of active play per team, excluding stoppages. & seconds \\
\hline
Goal difference & Total goals scored minus goals conceded. & count \\
\hline
\end{tabular}
\caption{\textbf{Supplementary Table S2.} Statistical performance metrics computed from match event data. Metrics are divided into those reported in the main analysis (top section) and additional measures computed for exploratory analysis (bottom section).}
\label{tab:SI_stat_metrics}
\end{table}

\clearpage

\subsection*{Network metrics}

In Table \ref{tab:SI_network_metrics}, we list the network measures analyzed in this work.

\begin{table}[ht]
\centering
\small
\begin{tabular}{|p{4cm}|p{7cm}|p{3cm}|}
\hline
\textbf{Metric} & \textbf{Description} & \textbf{Level} \\
\hline
\multicolumn{3}{|c|}{\textit{Reported in article}} \\
\hline
Outreach (mean) & Spatially weighted dispersion of passes; sum of pass counts multiplied by Euclidean distance between connected regions, normalized by total strength. Higher values indicate longer-range ball circulation. & Network \\
\hline
Maximum eigenvalue & Principal eigenvalue of weighted adjacency matrix; reflects overall connectivity and hierarchical structure. Higher values suggest more centralized passing patterns. & Network \\
\hline
Average shortest path & Mean minimum number of steps to connect any two regions (using inverse weights as distances). Lower values indicate more direct ball progression. & Network \\
\hline
\multicolumn{3}{|c|}{\textit{Additional metrics computed}} \\
\hline
In-strength & Total weight of incoming edges to a node; represents number of passes received by a pitch position. & Node (aggregated) \\
\hline
Out-strength & Total weight of outgoing edges from a node; represents number of passes originated from a pitch position. & Node (aggregated) \\
\hline
Betweenness centrality & Fraction of shortest paths passing through a node; identifies most commonly traversed pitch regions. & Node (aggregated) \\
\hline
Eigenvector centrality & Element of principal eigenvector; represents importance of pitch position within passing network. & Node (aggregated) \\
\hline
In/Out-closeness & Average distance from/to all other nodes; indicates accessibility of a pitch position for receiving/distributing the ball. & Node (aggregated) \\
\hline
Total strength & Sum of all edge weights; equivalent to total number of successful passes. & Network \\
\hline
Isolated nodes & Number of pitch regions with no incoming or outgoing passes. & Network \\
\hline
Self-loops & Passes that start and end in the same region (mean weight, max weight, count). & Network \\
\hline
Degree assortativity & Correlation between node degrees; indicates whether highly connected regions tend to connect to other highly connected regions (computed for in-in, out-out, in-out, out-in combinations). & Network \\
\hline
\end{tabular}
\caption{\textbf{Supplementary Table S3.} Network metrics computed from pitch-passing networks. Metrics are divided into those reported in the main analysis (top section) and additional measures computed for exploratory analysis (bottom section). Node-level measures were aggregated as mean and standard deviation across all 50 pitch regions.}
\label{tab:SI_network_metrics}
\end{table}

\clearpage

\section*{Additional performance metrics}

The following sections present detailed temporal and hierarchical patterns for supplementary performance indicators not included in the main manuscript. These metrics provide additional context for understanding tactical evolution across men's and women's football. Each subsection examines tier-specific trends and gender comparisons, complementing the core findings reported in the article.

\subsection*{Passing volume and shot buildup}

\begin{figure}[ht]
\centering
\includegraphics[width=0.8\linewidth]{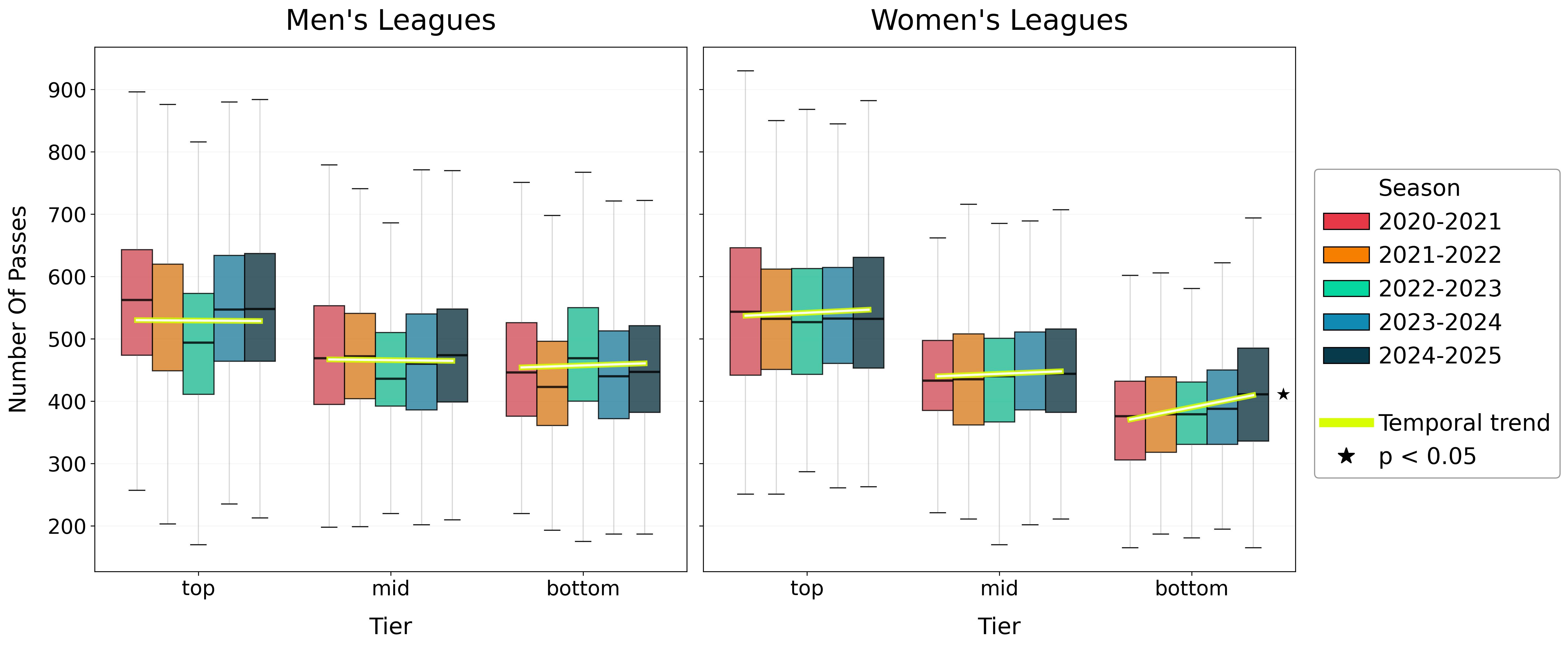}
\caption{\textbf{Supplementary Figure S2.} Total number of passes by tier for men's (left) and women's (right) football across five seasons. Top teams maintain the highest passing volumes across all seasons, with women's bottom-tier teams showing the only significant temporal increase.}
\label{fig:SI_passes}
\end{figure}

Figure~\ref{fig:SI_passes} presents the total number of passes completed by teams across tiers and seasons. Top-tier teams consistently complete the highest pass volumes (men: $\sim$541, women: $\sim$543), followed by mid-tier ($\sim$470 and $\sim$444) and bottom-tier teams ($\sim$451 and $\sim$391). Women's bottom-tier teams show a notable increasing pattern over time (slope = 8.69, $p = 0.04$), suggesting potential tactical development at lower competitive levels, though this trend is not observed in other tiers or in men's football.

\begin{figure}[ht]
\centering
\includegraphics[width=0.8\linewidth]{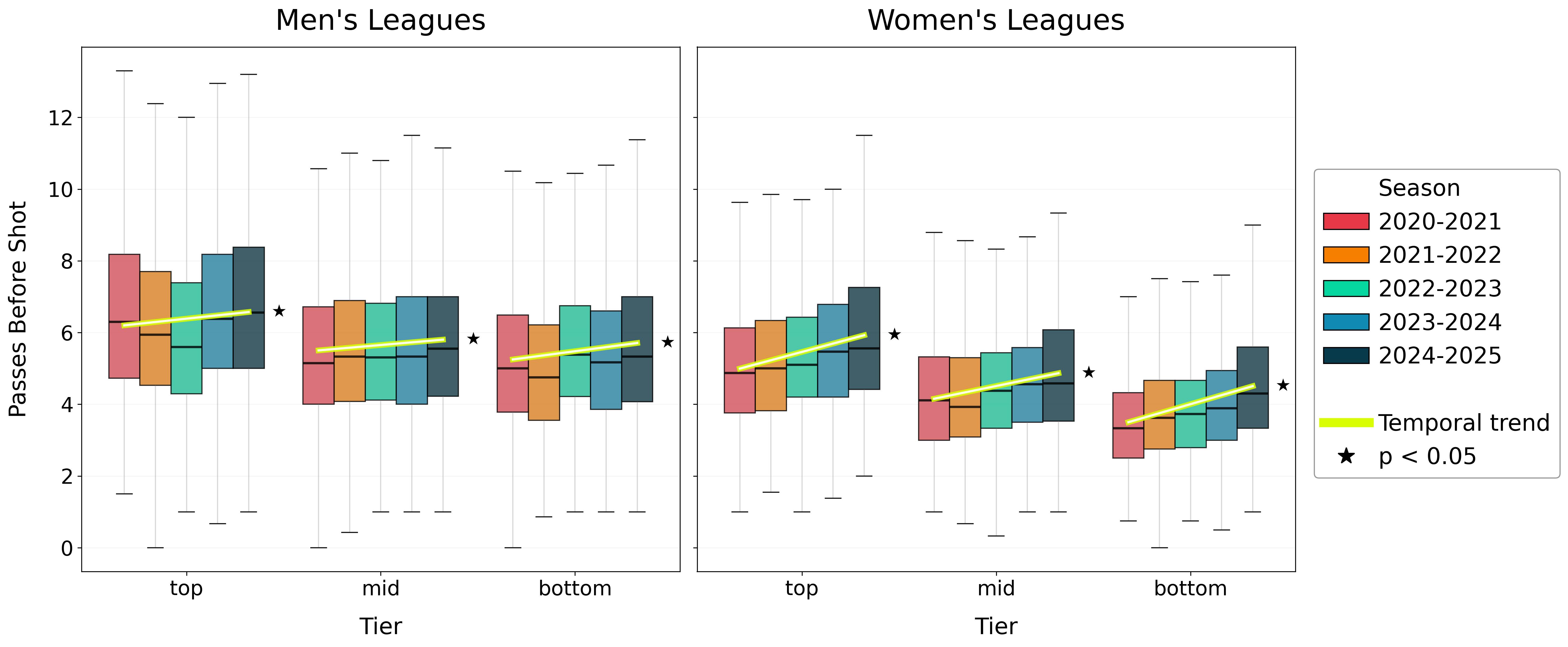}
\caption{\textbf{Supplementary Figure S3.} Passes before shot by tier for men's (left) and women's (right) football. All tiers exhibit increasing trends, with women's football showing significant growth across all competitive levels (all p < 0.05).}
\label{fig:SI_pbs}
\end{figure}

Passes before shot (PBS), quantifying buildup length prior to goal attempts, exhibits the expected performance hierarchy with top teams recording the highest values (Figure~\ref{fig:SI_pbs}). All tiers in both genders show increasing patterns over the five-year period, with women's football demonstrating particularly consistent growth across all competitive levels. This temporal trend complements the increases observed in passes per possession, collectively indicating a shift toward more deliberate attacking sequences.

\subsection*{Passing precision and shooting behavior}

\begin{figure}[ht]
\centering
\includegraphics[width=0.8\linewidth]{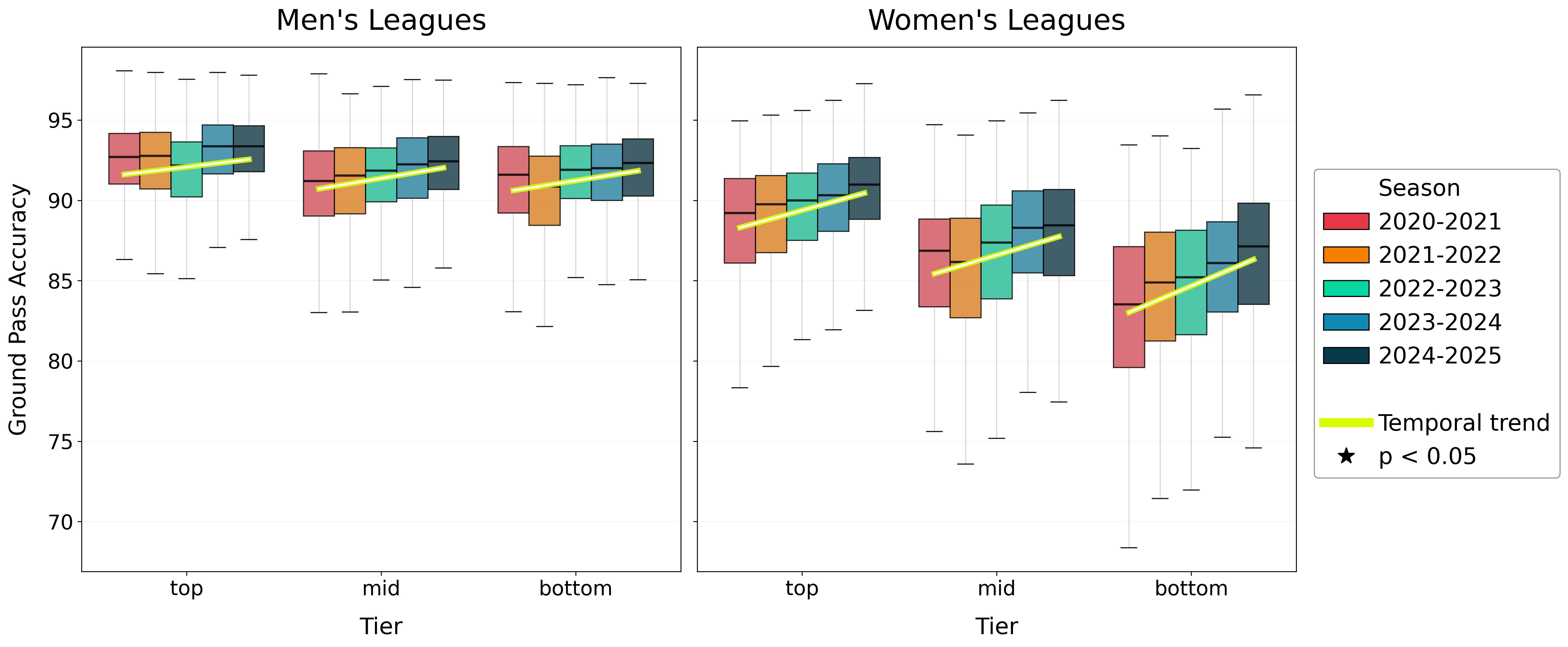}
\caption{\textbf{Supplementary Figure S4.} Ground pass accuracy by tier for men's (left) and women's (right) football. Women's competitions show stronger temporal increases than men's, with significant trends in top, mid (men only), and bottom tiers.}
\label{fig:SI_ground_acc}
\end{figure}

Ground pass accuracy---the completion rate of passes made at ground level---reveals systematic gender differences and temporal evolution (Figure~\ref{fig:SI_ground_acc}). Men's leagues average 90-95\% accuracy with modest increases over time, particularly in mid-tier teams. Women's leagues average 85-90\% with stronger positive trends, especially pronounced in top and bottom tiers. The gender gap ranges from approximately 3\% (top teams) to 8\% (bottom teams), consistent with overall passing accuracy patterns.

\begin{figure}[ht]
\centering
\includegraphics[width=0.8\linewidth]{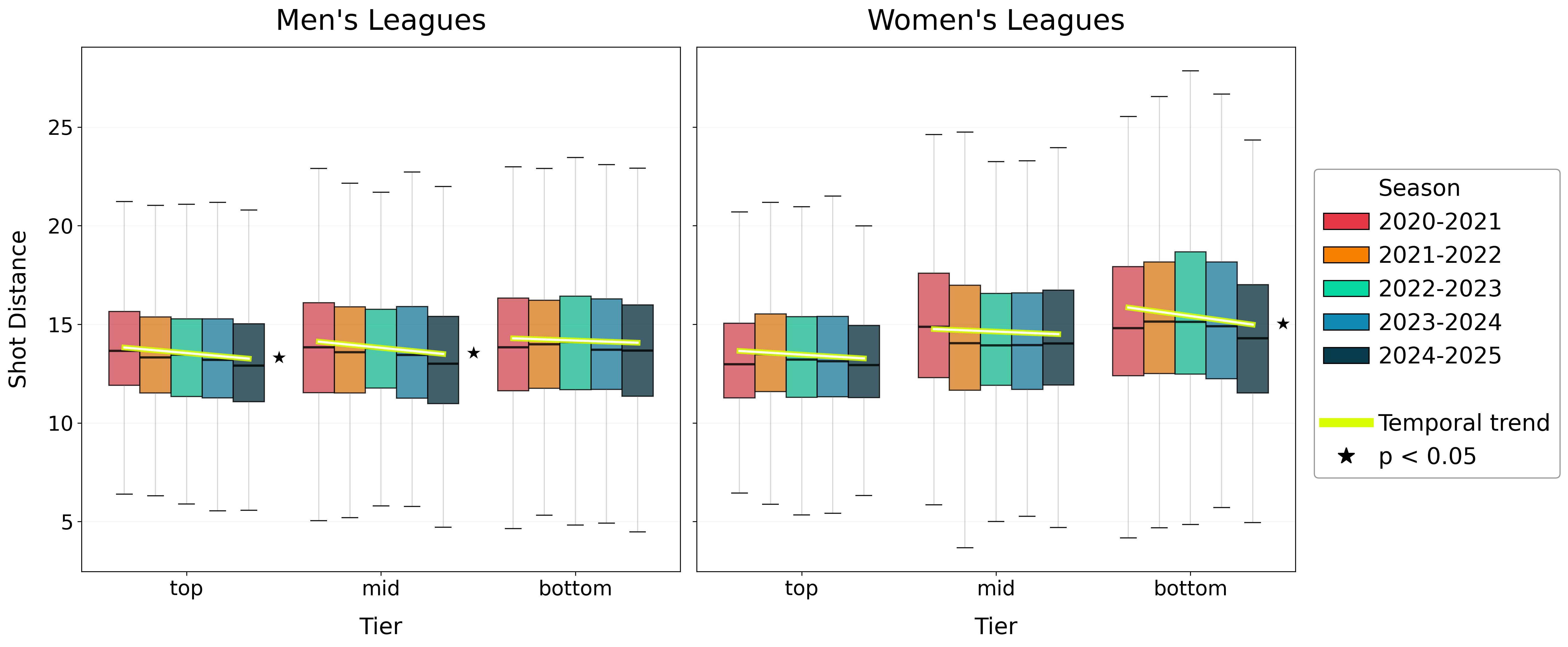}
\caption{\textbf{Supplementary Figure S5.} Average shot distance to goal by tier for men's (left) and women's (right) football. Men's top and mid-tier teams show significant declining trends, indicating progressively closer shooting positions over time.}
\label{fig:SI_shot_dist}
\end{figure}

Shot distance exhibits a consistent declining trend across most tiers in men's football (Figure~\ref{fig:SI_shot_dist}), with top and mid-tier teams showing significant decreases of approximately 0.14-0.15m per year. Women's football shows similar declining patterns, though not reaching statistical significance in most cases. These trends suggest teams are working the ball closer to goal before shooting, aligning with the observed increases in passes before shot and the shorter average pass lengths.

\begin{figure}[ht]
\centering
\includegraphics[width=0.8\linewidth]{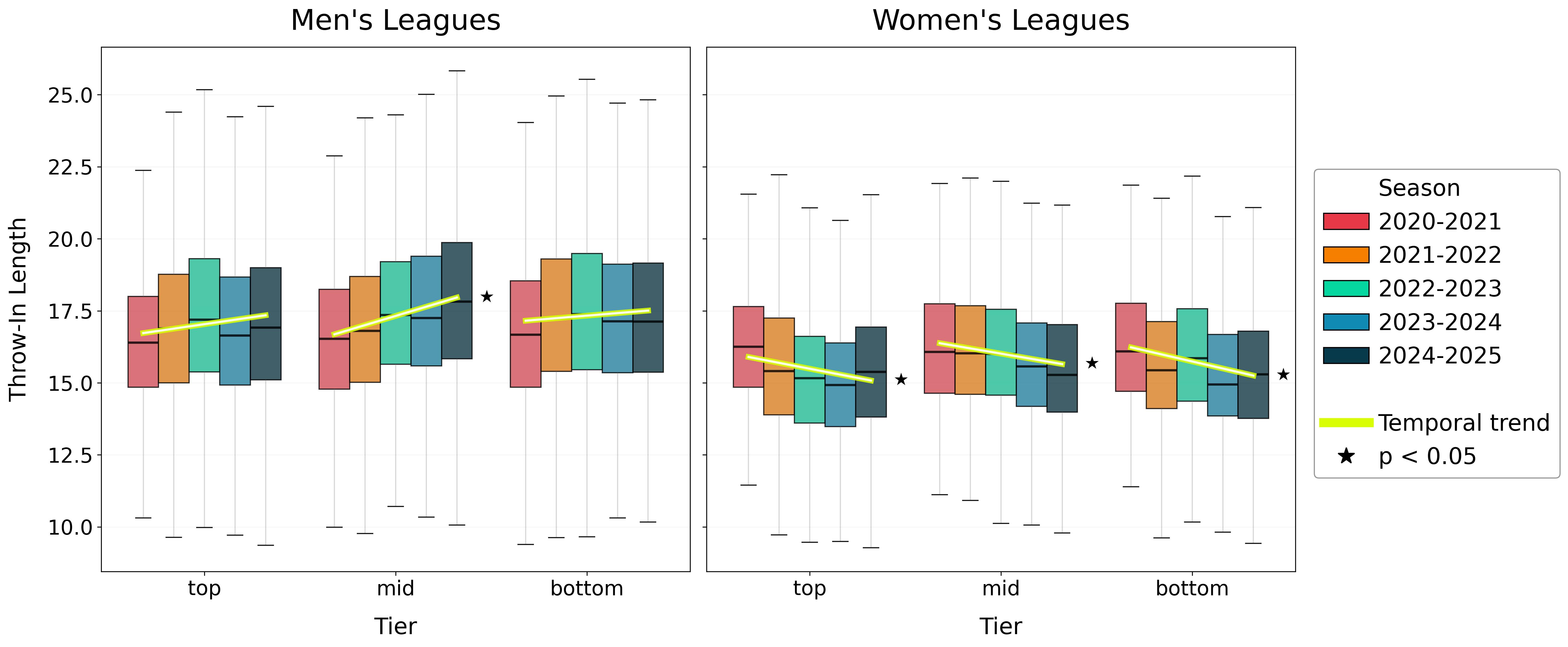}
\caption{\textbf{Supplementary Figure S6.} Average throw-in length by tier for men's (left) and women's (right) football. Men's leagues show stable or increasing distances, with mid-tier teams exhibiting a significant positive trend. Women's leagues display the opposite pattern, with decreasing distances and significant negative trends across all tiers.}
\label{fig:SI_throw-in_length}
\end{figure}

Throw-in length shows modest tier-based differences in both men's and women's leagues, with no consistent ranking pattern across top, mid, and bottom tiers (Figure~\ref{fig:SI_throw-in_length}). Temporal trends diverge sharply by gender. Men's leagues exhibit stable or increasing throw-in distances, with mid-tier teams showing a statistically significant positive trend. Women's leagues display the opposite pattern, with decreasing distances and significant negative trends across all tiers, suggesting gender-divergent tactical conventions around restart play.

\subsection*{Network structural properties}

\begin{figure}[ht]
\centering
\includegraphics[width=0.8\linewidth]{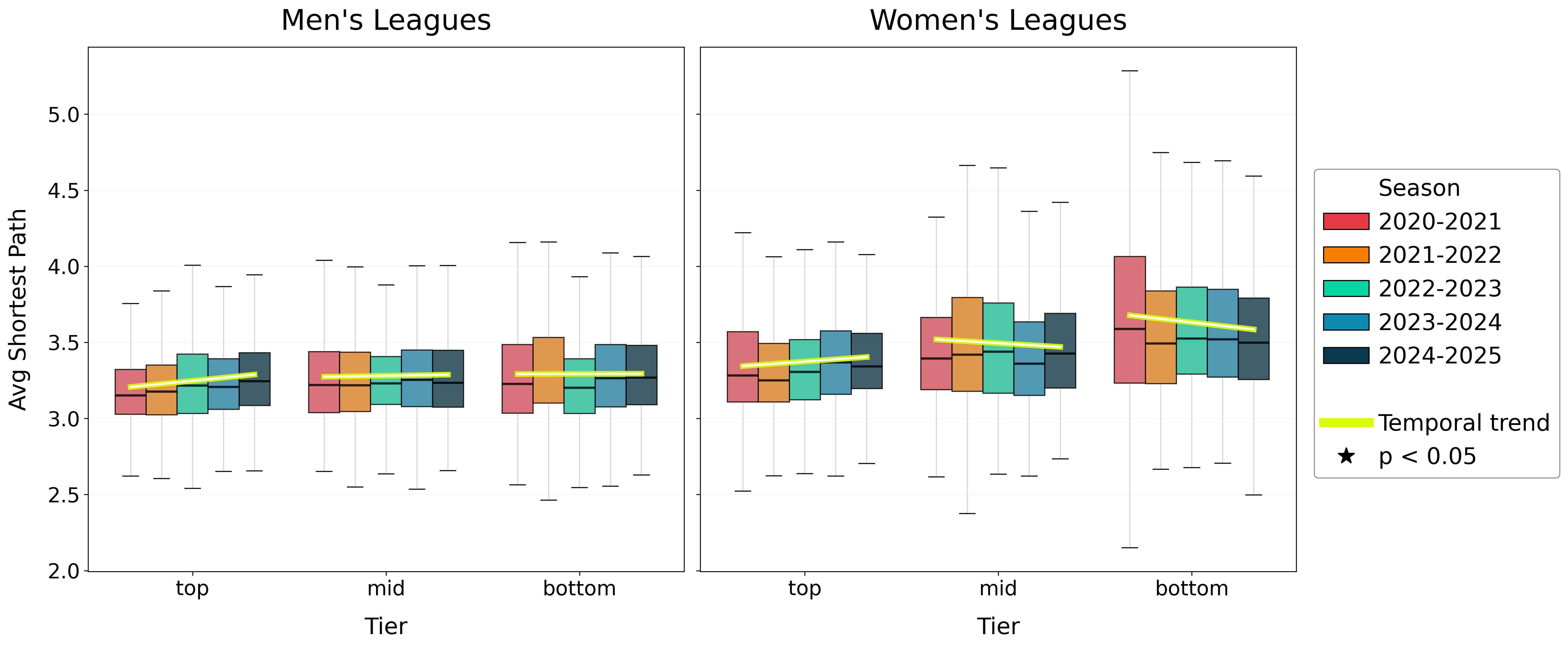}
\caption{\textbf{Supplementary Figure S7.} Average shortest path length in pitch-passing networks by tier for men's (left) and women's (right) football. Lower values indicate more efficient ball circulation. Men's top teams show a modest increasing pattern, though not statistically significant.}
\label{fig:SI_shortest_path}
\end{figure}

Average shortest path length quantifies the efficiency of ball circulation through the network (Figure~\ref{fig:SI_shortest_path}). Top teams maintain the shortest paths (~3.2-3.4), enabling rapid ball progression to any pitch region. Men's top-tier teams show a slight increasing pattern, while women's football exhibits relatively stable values across all tiers. The modest changes in this metric, combined with declining network outreach, suggest that teams are maintaining circulation efficiency while concentrating spatial distribution.

\clearpage

\section*{Statistical significance analysis}

The following tables provide comprehensive statistical summaries of all analyses presented throughout this study. Table~\ref{tab:SI_temporal} reports temporal trend coefficients for each metric across tiers and genders. Table~\ref{tab:SI_tier} presents within-tier comparisons, demonstrating systematic performance hierarchies. Table~\ref{tab:SI_gender} quantifies gender differences within each competitive tier. Significance levels are indicated by asterisks (*p < 0.05).

\begin{table}[ht]
\centering
\small
\begin{tabular}{|l|ccc|ccc|}
\hline
 & \multicolumn{3}{c|}{\textbf{Men}} & \multicolumn{3}{c|}{\textbf{Women}} \\
\textbf{Metric} & Top & Mid & Bottom & Top & Mid & Bottom \\
\hline
Number of passes & -1.20 & -0.77 & 1.67 & 1.37 & 1.85 & 8.69* \\
Passes per possession & 0.15 & 0.13* & 0.13 & 0.18* & 0.14* & 0.19* \\
Pass accuracy (\%) & 0.31 & 0.38 & 0.43 & 0.91* & 0.67 & 1.55* \\
Ground pass accuracy (\%) & 0.22 & 0.33* & 0.30 & 0.54* & 0.52 & 0.84* \\
Passes before shot & 0.08 & 0.07* & 0.11 & 0.23* & 0.17* & 0.24* \\
Passes under pressure & 3.52* & 2.75* & 2.71* & 3.79* & 2.83* & 3.68* \\
Shot distance (m) & -0.14* & -0.15* & -0.05 & -0.07 & -0.09 & -0.16 \\
Max eigenvalue & 0.03 & -0.02 & 0.05 & 0.34 & 0.12 & 0.36* \\
Avg shortest path & 0.02* & 0.00 & 0.00 & 0.01 & -0.01 & -0.03 \\
Network outreach & -0.18* & -0.06* & -0.15* & -0.36* & -0.25* & -0.40* \\
Vertical play & 0.01 & 0.01 & 0.01 & 0.02* & 0.02 & 0.04* \\
Throw-in length (m) & 0.16 & 0.32* & 0.09 & -0.25 & -0.17* & -0.26 \\
\hline
\end{tabular}
\caption{\textbf{Supplementary Table S4.} Temporal trends in match performance metrics (2020-2025). Values represent annual slope from linear regression. *p < 0.05.}
\label{tab:SI_temporal}
\end{table}

\begin{table}[ht]
\centering
\small
\begin{tabular}{|l|cccc|cccc|}
\hline
 & \multicolumn{4}{c|}{\textbf{Men}} & \multicolumn{4}{c|}{\textbf{Women}} \\
\textbf{Metric} & Top & Mid & Bottom & F & Top & Mid & Bottom & F \\
\hline
Number of passes & 541.1 & 469.6 & 451.4 & 32.6* & 542.9 & 444.1 & 391.1 & 293.6* \\
Passes per possession & 6.2 & 5.4 & 5.2 & 14.7* & 5.5 & 4.5 & 3.9 & 37.8* \\
Pass accuracy (\%) & 81.9 & 78.4 & 77.4 & 21.2* & 78.2 & 72.6 & 68.7 & 33.3* \\
Ground pass acc. (\%) & 92.3 & 91.4 & 91.2 & 5.4* & 89.4 & 86.6 & 84.6 & 24.0* \\
Passes before shot & 6.5 & 5.7 & 5.4 & 25.0* & 5.5 & 4.5 & 4.0 & 22.7* \\
Passes under pressure & 55.0 & 48.4 & 47.4 & 308.3* & 60.3 & 50.2 & 44.4 & 306.7* \\
Shot distance (m) & 13.5 & 13.8 & 14.1 & 9.3* & 13.4 & 14.7 & 15.4 & 56.4* \\
Max eigenvalue & 16.4 & 13.6 & 13.1 & 26.4* & 15.8 & 12.5 & 10.7 & 106.6* \\
Avg shortest path & 3.2 & 3.3 & 3.3 & 6.4* & 3.4 & 3.5 & 3.6 & 43.2* \\
Network outreach & 21.2 & 22.0 & 22.1 & 669.6* & 20.4 & 21.2 & 21.4 & 255.4* \\
Vertical play & 1.2 & 1.1 & 1.0 & 50.2* & 1.0 & 0.9 & 0.8 & 32.5* \\
Throw-in length (m) & 17.0 & 17.3 & 17.3 & 0.9 & 15.5 & 16.0 & 15.8 & 1.4 \\
\hline
\end{tabular}
\caption{\textbf{Supplementary Table S5.} Within-tier comparison of match performance metrics. Values represent tier means; F = ANOVA F-statistic. *p < 0.05.}
\label{tab:SI_tier}
\end{table}

\begin{table}[ht]
\centering
\small
\begin{tabular}{|l|ccc|ccc|ccc|}
\hline
 & \multicolumn{3}{c|}{\textbf{Top}} & \multicolumn{3}{c|}{\textbf{Mid}} & \multicolumn{3}{c|}{\textbf{Bottom}} \\
\textbf{Metric} & M & W & $\Delta$\% & M & W & $\Delta$\% & M & W & $\Delta$\% \\
\hline
Number of passes & 541.1 & 542.9 & -0.3 & 469.6 & 444.1 & +5.7* & 451.4 & 391.1 & +15.4* \\
Passes per poss. & 6.2 & 5.5 & +12.8* & 5.4 & 4.5 & +20.9* & 5.2 & 3.9 & +31.5* \\
Pass accuracy (\%) & 81.9 & 78.2 & +4.8* & 78.4 & 72.6 & +7.9* & 77.4 & 68.7 & +12.6* \\
Ground pass acc. (\%) & 92.3 & 89.4 & +3.3* & 91.4 & 86.6 & +5.4* & 91.2 & 84.6 & +7.7* \\
Passes before shot & 6.5 & 5.5 & +19.2* & 5.7 & 4.5 & +25.2* & 5.4 & 4.0 & +35.1* \\
Passes under pressure & 55.0 & 60.3 & -8.9* & 48.4 & 50.2 & -3.5* & 47.4 & 44.4 & +6.7* \\
Shot distance (m) & 13.5 & 13.4 & +0.6 & 13.8 & 14.7 & -5.9* & 14.1 & 15.4 & -8.1* \\
Max eigenvalue & 16.4 & 15.8 & +3.5 & 13.6 & 12.5 & +9.6* & 13.1 & 10.7 & +22.6* \\
Avg shortest path & 3.2 & 3.4 & -4.1* & 3.3 & 3.5 & -6.0* & 3.3 & 3.6 & -9.1* \\
Network outreach & 21.2 & 20.4 & +4.0* & 22.0 & 21.2 & +3.9* & 22.1 & 21.4 & +3.0* \\
Vertical play & 1.2 & 1.0 & +14.1* & 1.1 & 0.9 & +20.3* & 1.0 & 0.8 & +25.1* \\
Throw-in length (m) & 17.0 & 15.5 & +9.7* & 17.3 & 16.0 & +8.2* & 17.3 & 15.8 & +9.9* \\
\hline
\end{tabular}
\caption{\textbf{Supplementary Table S6.} Gender comparison of match performance metrics by tier. M = Men, W = Women, $\Delta$\% = percent difference (positive = men higher). *p < 0.05.}
\label{tab:SI_gender}
\end{table}

\subsection*{Offsides by country}

Offside infractions provide insight into tactical risk-taking and attacking positioning. Table~\ref{tab:SI_offsides} presents offside counts by country, revealing that Spain exhibits significantly elevated offside rates compared to all other leagues in both men's (+20\%) and women's football (+21\%). This pattern remained stable across the five-year period with no significant temporal trends observed.

\begin{table}[ht]
\centering
\begin{tabular}{|l|cc|c|}
\hline
\textbf{Country} & \textbf{Men} & \textbf{Women} & \textbf{Combined} \\
\hline
Spain & 2.06 & 2.54 & 2.22*** \\
Germany & 1.85 & 2.27 & 1.98 \\
Italy & 1.67 & 2.26 & 1.81 \\
England & 1.72 & 1.95 & 1.78 \\
USA & 1.65 & 1.93 & 1.71 \\
\hline
\textbf{Others (pooled)} & \textbf{1.71} & \textbf{2.10} & \textbf{1.81} \\
\hline
\end{tabular}
\caption{\textbf{Supplementary Table S7.} Offsides per match by country (2020-2025). Spain shows significantly higher offsides than all other countries pooled in both men's (2.06 vs 1.71, t=16.60, p<0.0001) and women's football (2.54 vs 2.10, p<0.0001), representing a +20\% and +21\% increase respectively. No significant temporal trends were observed (all p>0.05). ***p < 0.0001 vs pooled others.}
\label{tab:SI_offsides}
\end{table}

\subsection*{Spatial positioning: center of mass}

The x-coordinate of passing center of mass quantifies average attacking depth, indicating how high up the pitch teams position their buildup play. Table~\ref{tab:SI_center_mass} reveals that men's teams consistently position 6.5\% higher upfield than women's teams. Notably, Serie A Women exhibited a strong temporal increase in attacking positioning (+0.82 m/year), while other competitions showed minimal or opposite trends.

\begin{table}[ht]
\centering
\begin{tabular}{|l|cc|c|}
\hline
\textbf{Group} & \textbf{Mean} & \textbf{Temporal Trend} & \textbf{Significance} \\
 & \textbf{(m)} & \textbf{(m/year)} & \\
\hline
All Men & 55.71 & $-0.12$ & *** \\
All Women (excl. Italy) & 52.30 & $+0.15$ & * \\
\textbf{Serie A Women} & \textbf{52.33} & \textbf{+0.82} & \textbf{***} \\
\hline
\end{tabular}
\caption{\textbf{Supplementary Table S8.} Center of mass (x-coordinate) by group (2020-2025). Serie A Women show a strong temporal increase in attacking positioning (+0.82 m/year, p<0.0001), while other groups exhibit minimal or opposite trends (All Men: $-0.12$ m/year, p<0.0001; Women excl. Italy: +0.15 m/year, p=0.03). Men's teams position significantly higher upfield than women's (+6.5\%, p<0.0001), but no difference exists between women's groups (p=0.89). *p < 0.05, ***p < 0.0001.}
\label{tab:SI_center_mass}
\end{table}

\end{document}